\documentclass[aps,prd,onecolumn,eqsecnum,amsmath,nofootinbib,preprintnumbers]{revtex4}%

\usepackage{color,graphicx,float,subfigure,xcolor}
\usepackage{amsfonts,amssymb,theorem,mathrsfs,times}
\usepackage{bm}
\usepackage{mathtools}
\usepackage{amsfonts,amssymb,theorem,mathrsfs}
\usepackage{dsfont}
\usepackage{setspace}
\usepackage{amsmath}  
\usepackage[colorlinks,linkcolor=blue,anchorcolor=blue,citecolor=blue]{hyperref}   
\usepackage{ulem}   
\usepackage[english]{babel}
\usepackage{tikz}

\usepackage{accents}
\usepackage{url}

{\theorembodyfont{\upshape}
	}
{\theorembodyfont{\upshape}
	}
{\theorembodyfont{\upshape}
	}
{\theorembodyfont{\upshape}
	}
{\theorembodyfont{\upshape}
	}
{\theorembodyfont{\upshape}
	}
\addtocounter{MaxMatrixCols}{10}
\newcommand{\dalm}{\kern1pt\vbox{\hrule height 0.9pt\hbox{\vrule width
			0.9pt\hskip 2.5pt\vbox{\vskip 5.5pt}\hskip 3pt\vrule width
			0.3pt}\hrule height 0.3pt}\kern1pt}

\begin{document}
\thispagestyle{empty}
\title{The parameterized quasinormal modes for modified Teukolsky equations}
	
%

\author{Zhe Yu$^{a}$\footnote{e-mail address: yuzhe@nbu.edu.cn}} 




\author{Liang-Bi Wu$^{b\, ,c}$\footnote{e-mail address: liangbi@mail.ustc.edu.cn (corresponding author)} }
 
	
\affiliation{${}^a$Institute of Fundamental Physics and Quantum Technology, Ningbo University, Ningbo 315211, China}

\affiliation{${}^b$School of Fundamental Physics and Mathematical Sciences, Hangzhou Institute for Advanced Study, UCAS, Hangzhou 310024, China}


\affiliation{${}^c$University of Chinese Academy of Sciences, Beijing 100049, China}



\date{\today}
	
\begin{abstract}
We introduce the modified Teukolsky equation within a parameterized framework, analogous to the case of small deviations of potential in spherical symmetry. Both the radial and angular equations acquire modifications described by two independent sets of parameters. We derive the parameterized framework of the quasinormal mode spectra using the continued fraction method. The results are cross-validated with the two-dimensional pseudo-spectral method, demonstrating excellent agreement and ensuring self-consistency. This work establishes a robust foundation for a theory-agnostic interpretation of gravitational-wave ringdown signals, providing a practical tool for probing potential deviations from General Relativity in the strong-field regime.
\end{abstract}
	
\maketitle
	
\section{Introduction}
Gravitational-wave (GW) astronomy has made tremendous strides in recent years. The detection of over a hundred binary black hole (BH) mergers has enabled precise tests of general relativity (GR) in the strong-field regime~\cite{KAGRA:2021vkt}. In these events, the merger produces a highly distorted black hole, which subsequently settles into a stationary black hole through the emission of GWs. This process, known as the ringdown phase, is characterized by a damped oscillatory signal that can be decomposed into quasinormal modes (QNMs)~\cite{Nollert:1999ji,Kokkotas:1999bd,Berti:2025hly}. During the ringdown, the perturbations are sufficiently small, allowing the linear approximation to hold. In the Kerr spacetime, QNMs are solutions to the Teukolsky equation (TE) that satisfy the radiation boundary condition which have been successfully applied to interpret the ringdown signals of dozens of observed events~\cite{LIGOScientific:2016lio,LIGOScientific:2020tif,LIGOScientific:2021sio}.

The primary scientific driver for a majority of GW detection projects is the search for deviations from GR, which could originate from modified theories of gravity or other new physical scenarios~\cite{LIGOScientific:2020tif,LIGOScientific:2021sio,Gong:2021gvw,Bigongiari:2025oyk}. In ringdown phase, a significant challenge in this endeavor is the direct computation of QNMs when the fundamental theory of gravity or its corresponding black hole solutions remain unspecified. In response, a theory-agnostic, parametric framework for calculating QNMs has been rapidly developed in recent years~\cite{Berti:2025hly}, enabling the systematic investigation of a rich phenomenology that captures various classes of modifications through a set of deviation parameters. This approach effectively balances theoretical generality with practical simplicity, allowing efficient computation of modified QNMs that prove invaluable for data analysis aimed at discovering new physics~\cite{Gossan:2011ha,Meidam:2014jpa,Carullo:2018sfu,LIGOScientific:2020tif,LIGOScientific:2021sio}. The parameterization must incorporate constraints to remain physically meaningful, drawing partly from stringent experimental bounds that indicate any deviations from GR must be minimal~\cite{LIGOScientific:2018dkp,LIGOScientific:2019fpa,LIGOScientific:2021sio,Yunes:2024lzm}, and partly from theoretical requirements, such as the ansatz that modifications to the QNMs originate from the deformation of the effective potential in the master equation governing single-field perturbations in spherically symmetric spacetime. This assumption inherently demands the separability of the perturbation equations, consequently restricting the framework to specific classes of beyond-GR theories~\cite{Tattersall:2019nmh,deRham:2020ejn,Roussille:2022vfa,Sirera:2023pbs,Mukohyama:2023xyf}. More concretely, the methodology typically employs linear parametrization of potential deviations, with the modified equations solved via the continued fraction method~\cite{Leaver:1985ax} to yield QNM spectra at first order in the small parameters. This framework has been successfully implemented in spherically symmetric spacetimes~\cite{Cardoso:2019mqo,Hatsuda:2020egs,Volkel:2022aca}, with important extensions including quadratic corrections~\cite{McManus:2019ulj}, QNM overtone calculations~\cite{Hirano:2024fgp}, and time-domain evolution studies~\cite{Thomopoulos:2025nuf}. A crucial recent advance generalizes this approach to rotating spacetimes by solving a modified TE through variable separation. While the angular equation retains its standard form, the radial equation incorporates modifications analogous to the potential deformation in the spherical case. The key distinction from the spherical case lies in the requirement to determine both the separation constant and the QNM spectra simultaneously through the deviation parameters~\cite{Cano:2024jkd}.

Besides the semi-analytic method within this parameterized framework, the pseudo-spectral method emerges as a powerful and complementary numerical tool~\cite{Jaramillo:2020tuu,Cai:2025irl,Jansen:2017oag,Ripley:2022ypi,PanossoMacedo:2024nkw,Xiong:2024urw}. This approach discretizes the differential eigenvalue problem directly on a spatial grid, effectively converting it into a large-scale matrix problem whose eigenvalues yield the QNM spectra. Its principal advantage lies in its ability to handle the full two-dimensional nature of perturbations in rotating spacetimes natively, bypassing the need for a separable ansatz that underpins many traditional methods~\cite{Blazquez-Salcedo:2023hwg,Chung:2023wkd}. Consequently, the pseudo-spectral method offers a highly accurate and computationally efficient pathway to calculate QNMs, not only within GR but also for a broader class of modified gravity theories~\cite{Blazquez-Salcedo:2024oek,Chung:2024ira,Chung:2024vaf}. This makes it an invaluable asset for cross-validating results from the parametric framework and for exploring a wider, less constrained theory space in the ongoing search for deviations from GR.

In this work, we analyze a modified Teukolsky equation that incorporates an additional term proportional to both the perturbation field and a set of deviation parameters, following the spirit of master equations in spherically symmetric settings, namely modification of the effective potential~\cite{Cheung:2021bol,Berti:2022xfj,Courty:2023rxk,Yang:2024vor,Cardoso:2024mrw,Ianniccari:2024ysv,MalatoCorrea:2025iuc,Shen:2025nsl,Mai:2025cva}. Within our parameterization, both modifications appear in the angular and radial equations, obtained after separation of variables. Accordingly, the final parameterized QNM spectra are expressed in terms of two independent sets of parameters. To constrain the parameters in this setup and ensure the consistency of the results at first order in the deviation parameters, the pseudo-spectral method is employed for cross-validation. Overall, this work establishes a foundation for developing a theory-informed, parametric description of QNMs in the context of generic, agnostic deviations from GR within the context of a two-dimensional spectral problem.

This paper is organized as follows. In Sec.~\ref{sec:TE_M}, we present the modified Teukolsky equation, and show the resulting modified radial and angular equation. Sec.~\ref{Sec: Parameterized_framework} details the application of the continued fraction method to solve the resulting system and expresses QNM spectra in terms of the deviation parameters linearly to achieve the parameterized framework. In Sec.~\ref{Sec: QNM_two_dimensioncal}, we employ the pseudo-spectral method for independent cross-validation and derive constraints on the parameters. Conclusions and broader discussions are provided in Sec.~\ref{sec:Conlusion}. Furthermore, Appendix~\ref{app:1} presents the detailed derivation of recurrence relations within the continued fraction method, while Appendix~\ref{app:2} provides the explicit form of the operator $L$ used in the pseudo-spectral setup. Throughout this work, we use the mostly plus metric signature and adopt geometric units with $c = G = 1$.

\section{Modified Teukolsky equations}
\label{sec:TE_M}
We begin this section by reviewing the metric of the Kerr black hole in the Boyer-Lindquist (BL) coordinates $(t,r,\theta,\varphi)$
\begin{eqnarray}\label{Kerr_metric_BL}
    \mathrm{d}s^2=-\Big(1-\frac{2Mr}{\Sigma}\Big)\mathrm{d}t^2-\frac{4Mar}{\Sigma}\sin^2\theta\mathrm{d}t\mathrm{d}\varphi+\frac{\Sigma}{\Delta}\mathrm{d}r^2+\Sigma\mathrm{d}\theta^2+\sin^2\theta\Big(\Sigma_0+\frac{2Ma^2r}{\Sigma}\sin^2\theta\Big)\mathrm{d}\varphi^2\, ,
\end{eqnarray}
where
\begin{eqnarray}
    \Delta(r)&=&r^2-2Mr+a^2=(r-r_{+})(r-r_{-})\, ,\nonumber\\
    \Sigma(r,\theta)&=&r^2+a^2\cos^2\theta\, ,\quad \Sigma_0(r)=\Sigma(r,0)=r^2+a^2\, ,
\end{eqnarray}
here $M$ and $a$ denote the mass and angular momentum parameters of the black hole. Throughout this work, we set $M = 1/2$ (Although we will still carry $M$ in the expressions of subsequent contents.) and restrict the angular momentum parameter to the range $a \in [0, 1/2)$. The zeros of the function $\Delta$ determine the location of the BH Cauchy horizon and event horizon, which are given by
\begin{eqnarray}\label{event_horizon_and_Cauchy_horizon}
    r_{\pm}=M\pm\sqrt{M^2-a^2}=M(1\pm\beta)\, ,\quad \beta\equiv\sqrt{1-\Big(\frac{a}{M}\Big)^2}\, ,
\end{eqnarray}
where $r_{+}$ and $r_{-}$ are denoted as the event horizon and the Cauchy horizon of Kerr black hole, respectively.
The tortoise coordinate $r_\star$ for the Kerr black hole is given by
\begin{eqnarray}\label{tortoise_coordinate}
    r_{\star}=r+\frac{a^2+r_{+}^2}{r_{+}-r_{-}}\ln(r-r_{+})-\frac{a^2+r_{-}^2}{r_{+}-r_{-}}\ln(r-r_{-})\, .
\end{eqnarray}
Following the approach of modifying the master equation in the spherically symmetric case~\cite{Cheung:2021bol,Berti:2022xfj,Courty:2023rxk,Yang:2024vor,Cardoso:2024mrw,Ianniccari:2024ysv,MalatoCorrea:2025iuc,Shen:2025nsl,Mai:2025cva}, where deviations from modified gravity or new physical scenarios remain small, the background geometry can be maintained as in GR case while only the potential acquires a small modification. Given the above, to extend the parametrized framework for potential deviations from spherical symmetry to axisymmetric backgrounds, we introduce a generic perturbation term into the TE, where the modified equation takes the following form, 
\begin{eqnarray}\label{TE_with_perturbation}
    \Big[(\nabla^a+s\Gamma^a)(\nabla_a+s\Gamma_a)-4s^2\Psi_2\Big]\Psi^{(s)}=\frac{\eta}{\lambda^2} \Psi^{(s)}\, ,
\end{eqnarray}
where the function $\eta(r,\theta)$ encodes the deviation parameters with the angular dependence, while $\lambda$ serves as the characteristic length scale of the black hole, ensuring that $\eta$ remains dimensionless. In addition, $\Gamma^a$ denotes the connection vector, and $\Psi_2$ represents the nonvanished Weyl scalar in Kerr black hole~\cite{Bini:2002jx}. In the B-L coordinates, Eq. (\ref{TE_with_perturbation}) can be written as
\begin{eqnarray}\label{BL_TE_with_perturbation}
    0&=&\Big[\frac{(\Sigma_0)^2}{\Delta}-a^2\sin^2\theta\Big]\partial^2_{tt}\Psi^{(s)}+\frac{4Mar}{\Delta}\partial^2_{t\varphi}\Psi^{(s)}+\Big(\frac{a^2}{\Delta}-\frac{1}{\sin^2\theta}\Big)\partial^2_{\varphi\varphi}\Psi^{(s)}\nonumber\\
    &&-\Delta^{-s}\partial_r\Big(\Delta^{s+1}\partial_r\Psi^{(s)}\Big)-2s\Big[\frac{M(r^2-a^2)}{\Delta}-(r+\mathrm{i}a\cos\theta)\Big]\partial_t\Psi^{(s)}\nonumber\\
    &&-2s\Big[\frac{a(r-M)}{\Delta}+\mathrm{i}\frac{\cos\theta}{\sin^2\theta}\Big]\partial_{\varphi}\Psi^{(s)}-\frac{1}{\sin\theta}\partial_{\theta}\Big(\sin\theta\partial_\theta\Psi^{(s)}\Big)+s(s\cot^2\theta-1)\Psi^{(s)}\nonumber\\
    &&+(r^2+a^2\cos^2\theta)\frac{\eta(r,\theta)}{\lambda^2}\Psi^{(s)}\, .
\end{eqnarray}
To solve the above equation within the parameterized framework, we assume separability, requiring that $\eta(r,\theta)$ admits the decomposition
\begin{eqnarray}
\label{ansatz_separation}
    \frac{r^2+a^2\cos^2\theta}{\lambda^2}\eta(r,\theta)=\eta_1(r)+\eta_2(\theta)\, ,
\end{eqnarray}
where $\eta_1(r)$ and $\eta_2(\theta)$ are dimensionless functions. Naturally, the perturbation field $\Psi^{(s)}$ takes the separable form, namely
\begin{eqnarray}\label{separation_of_variables_Psi_s}
    \Psi^{(s)}(t,r,\theta,\varphi)=\mathrm{e}^{-\mathrm{i}\omega t}\mathrm{e}^{\mathrm{i}m\varphi}S(\theta)R(r)\, .
\end{eqnarray}
Substituting the above expression into the modified TE \eqref{BL_TE_with_perturbation} leads to its decoupling into separate radial and angular equations. On the one hand, the radial function $R(r)$ satisfies
\begin{eqnarray}\label{equation_R}
    \Delta^{-s}\frac{\mathrm{d}}{\mathrm{d}r}\Big[\Delta^{s+1}\frac{\mathrm{d}R(r)}{\mathrm{d}r}\Big]+\Big[V(r)-\eta_1(r)\Big]R(r)=0\, ,
\end{eqnarray}
where $V(r)$ is the potential in GR case. The potential $V(r)$ has the following form
\begin{eqnarray}\label{V_and_K}
    V(r)=\frac{K^2-2\mathrm{i}s(r-M)K}{\Delta}+4\mathrm{i}s\omega r-\Big(A_{l m}^{(s)}(a\omega)+a^2\omega^2-2am\omega\Big)\, ,\quad K\equiv (r^2+a^2)\omega-am\, ,
\end{eqnarray}
where $A_{l m}^{(s)}(a\omega)$ denotes the separation constant. The function $\eta_1(r)$ can be regard as a modified term for the original radial equation. On the other hand, the angular function $S(x)$, where $x=\cos\theta$, $\eta_2(x)\equiv\eta_2(\arccos x)$, and $S(x)\equiv S(\arccos x)$, satisfies
\begin{eqnarray}\label{equation_S}
    \frac{\mathrm{d}}{\mathrm{d}x}\Big[(1-x^2)\frac{\mathrm{d}S(x)}{\mathrm{d}x}\Big]+\Big[a^2\omega^2x^2-2a\omega sx+s+A_{l m}^{(s)}(a\omega)-\frac{(m+sx)^2}{1-x^2}-\eta_2(x)\Big]S(x)=0\, ,
\end{eqnarray}
where the function $\eta_2(x)$ acts as a modification to the original angular equation. In the following section, we will use Eq.~\eqref{equation_R} and Eq.~\eqref{equation_S} to study the parameterized framework for the modified Teukolsky equation.



\section{Parameterized framework of QNMs}\label{Sec: Parameterized_framework}
In this section, the radial Eq. \eqref{equation_R} and angular Eq. \eqref{equation_S} will be solved by continued fraction method in the parameterized framework. We begin with the angular sector, where the analysis is more tractable. The QNM condition requires the angular function $S(x)$ to be regular over $x \in [-1,1]$. To preserve this regularity while avoiding additional singularities in the equation, we restrict the deviation function $\eta_2(x)$ to be real and parameterize it through a finite polynomial expansion
\begin{eqnarray}\label{eta_2}
    \eta_2(x)=\sum_{k=0}^{K_2}\eta_2^{(k)}(1+x)^k\, ,
\end{eqnarray}
where $K_2$ is the maximum expansion order which is dependent on the specific model, and the summation starts at $k=0$ to maintain regularity at $x=-1$. The coefficients $\eta_2^{(k)}$ are treated as the adjustable deviation parameters. This parameterization permits the angular function $S(x)$ to be finite at the regular singular points $x=\pm1$ like the Kerr case~\cite{Leaver:1985ax}. The angular function $S(x)$ therefore takes the form
\begin{eqnarray}\label{ansatz_S}
    S(x)=(1+x)^{k_1}(1-x)^{k_2}\mathrm{e}^{a\omega x}\sum_{n=0}^{\infty}S_n(1+x)^n\, ,
\end{eqnarray}
in which $k_1=|m-s|/2$ and $k_2=|m+s|/2$. Substituting the solution ansatz (\ref{ansatz_S}) into Eq. \eqref{equation_S}, after some algebraic manipulation, we arrive at the recurrence relation governing the series coefficients 
\begin{eqnarray}\label{recurrence_angular}
    \epsilon[n-K_2]\mathbf{\Theta}^{(K_2)}_n S_{n-K_2}+\cdots+\epsilon[n-2]\mathbf{\Theta}^{(2)}_nS_{n-2}+\epsilon[n-1]\mathbf{\Theta}^{(1)}_nS_{n-1}+\epsilon[n]\mathbf{\Theta}^{(0)}_nS_n+\epsilon[n]\mathbf{\Theta}^{(-1)}_nS_{n+1}=0\, ,
\end{eqnarray}
with coefficients explicitly given by
\begin{eqnarray}
    \mathbf{\Theta}^{(-1)}_n&=&-4(n+1)\left(|m-s|+1+n\right)\, ,\nonumber\\
    \mathbf{\Theta}^{(0)}_n&=&2n^2+n\left(-8a\omega +2|m-s|+2|m+s|+2\right)-2a^2\omega^2-4a\omega\left(|m-s|+s+1\right)\nonumber\\
    &&+|m-s|+|m-s||m+s|+|m+s|-2A_{lm}^{(s)}(a\omega)+m^2-s^2-2s+2\eta_2^{(0)}\, ,\nonumber\\
    \mathbf{\Theta}^{(1)}_n&=&2a\omega\left( |m-s|+  |m+s|+2n+2s\right) +2\eta_2^{(1)}\, ,\nonumber\\
    \mathbf{\Theta}^{(2)}_n&=&2\eta_2^{(2)}\, ,\nonumber\\
    &\cdots&\nonumber\\
    \mathbf{\Theta}^{(K_2)}_n&=&2\eta_2^{(K_2)}\, ,\nonumber
\end{eqnarray}
where $\epsilon[n]$ $(n\in\mathbb{Z})$ is discrete unit step function with $\epsilon=1$ for $n\ge0$, and $\epsilon=0$ for $n<0$. Note that when all parameters $\eta_2^{(k)}$ vanish, the recurrence relation~\eqref{recurrence_angular} will turn back to the standard three-term recurrence relation in GR~\cite{Leaver:1985ax}.

Now we turn to consider the radial equation~\eqref{equation_R}. Consistent with the treatment of the angular case, the function $\eta_1(r)$ must remain finite at both boundaries to avoid introducing additional singularities in the equation governing local perturbations. This regularity condition requires that
\begin{eqnarray}
    \lim_{r\to r_{+}}\eta_1(r)=\text{finite}\, ,\quad\text{and}\quad \lim_{r\to \infty}\eta_1(r)=\text{finite}\, .
\end{eqnarray}
Following the above requirements, we can parameterize $\eta_1(r)$ as
\begin{eqnarray}\label{eta_1}
    \eta_1(r)=\sum_{k=0}^{K_1}\eta_1^{(k)}\Big(\frac{r_{+}-r_{-}}{r-r_{-}}\Big)^k\, ,
\end{eqnarray}
which maintains generality while satisfying the regularity condition, and $K_1$ is the maximum expansion order determined by the specific model. With this parameterization established, we now analyze the solution structure of Eq.~\eqref{equation_R} (details refer to Appendix~\ref{app:1}) and adopt the following ansatz
\begin{eqnarray}\label{R_Structure}
    R(r)=\Big(\frac{r-r_{+}}{r-r_{-}}\Big)^{-\mathrm{i}\sigma-s}(r-r_{-})^{p-2s-1}\mathrm{e}^{\mathrm{i}\omega r}\sum_{n=0}^{\infty}R_n\Big(\frac{r-r_{+}}{r-r_{-}}\Big)^n\, ,
\end{eqnarray}
in which the parameters $\sigma$ and $p$ are given by
\begin{eqnarray}
    \sigma\equiv\frac{K_{+}}{2\beta M}\, ,\quad K_{+}\equiv 2\omega Mr_{+}-am\, ,\quad p\equiv2\mathrm{i}M\omega\,  .
\end{eqnarray}
Substituting the solution ansatz (\ref{R_Structure}) into Eq. \eqref{equation_R}, after some algebraic manipulation, we arrive at the recurrence relation of the series coefficients $R_n$ which are satisfied with 
\begin{eqnarray}\label{recurrence_radial}
    \epsilon[n-K_1]\mathbf{R}^{(K_1)}_n R_{n-K_2}+\cdots+\epsilon[n-2]\mathbf{R}^{(2)}_nR_{n-2}+\epsilon[n-1]\mathbf{R}^{(1)}_nR_{n-1}+\epsilon[n]\mathbf{R}^{(0)}_nR_n+\epsilon[n]\mathbf{R}^{(-1)}_nR_{n+1}=0\, ,
\end{eqnarray}
where the coefficients have the following expressions
\begin{eqnarray}
    \mathbf{R}^{(-1)}_n&=&(n+1)(n-s-2\mathrm{i}\sigma +1)\, ,\nonumber\\
    \mathbf{R}^{(0)}_n&=&\omega^2\Big(\beta^2M^2+4\beta M^2-\frac{2M^2}{\beta^2}+5M^2\Big)+2\omega\Big(\frac{am}{\beta^2}+2\beta M\sigma+2 M \sigma+\mathrm{i}\beta M   + \mathrm{i} M\Big)\nonumber\\
    &&-2n^2+2n\Big[2\mathrm{i}\sigma+2\mathrm{i}(\beta +1) M \omega  -1\Big] -\frac{a^2 m^2}{2 \beta^2M^2}-A_{lm}^{(s)}(a\omega)+2\sigma^2+2\mathrm{i}\sigma-s-1-\sum_{j=0}^{K_1}\eta_1^{(j)}\nonumber\, ,\\
    \mathbf{R}^{(1)}_n&=&\omega^2\Big(\frac{M^2}{\beta ^2}-\frac{2M^2}{\beta} -3M^2\Big)+\omega\Big(-\frac{am}{\beta^2}+\frac{am}{\beta}  -4M\sigma+\frac{\mathrm{i}sM}{\beta}-3 \mathrm{i} M s \Big)+n^2+n(-4\mathrm{i}M \omega+s-2\mathrm{i}\sigma)\nonumber\\
    &&+\frac{a^2 m^2}{4 \beta ^2 M^2} -\frac{\mathrm{i}ams}{2 \beta  M}-\sigma ^2-\mathrm{i} s \sigma+\sum_{j=1}^{K_1}j\eta_1^{(j)}\, ,\nonumber\\
    \mathbf{R}^{(2)}_n&=&-\sum_{j=2}^{K_1}\binom{j}{2}\eta_1^{(j)}\, ,\nonumber\\
    &\cdots&\nonumber\\
    \mathbf{R}^{(k)}_n&=&(-1)^{k+1}\sum_{j=k}^{K_1}\binom{j}{k}\eta_1^{(j)}\, ,\nonumber\\
    &\cdots&\nonumber\\
    \mathbf{R}^{(K_1)}_n&=&(-1)^{K_1+1}\eta_1^{(K_1)}\, .\nonumber
\end{eqnarray}
Similarly, this recurrence relation will turn back to be the one in GR~\cite{Leaver:1985ax} when the deviation parameters $\eta_1^{(k)}(r)$ vanish totally.

It is known that the recurrence relations~\eqref{recurrence_angular} and~\eqref{recurrence_radial} can be reduced to three-term recurrence relations using Gaussian elimination~\cite{Volkel:2022aca,Konoplya:2011qq}, enabling the solutions for $S_n$ and $R_n$ to be expressed as continued fractions. However, achieving convergence requires truncating these continued fractions at large $n$, which in turn necessitates extensive Gaussian elimination operations. To avoid this computationally intensive process, we employ an alternative method that achieves equivalent results through a more efficient computational pathway. It can be seen that the recurrence relations form an infinite system of linear equations for the expansion coefficients $S_n$ and $R_n$. We truncate this system at a sufficiently large index $n$, reducing the problem to solving a finite homogeneous system. The existence of nontrivial solutions requires the determinant of the coefficient matrix to vanish. This condition yields coupled algebraic equations
\begin{eqnarray}\label{algebratic_Eq}                               \mathcal{L}_r(\omega,A,\eta_1^{(k)})=0\, ,\quad                 \mathcal{L}_\theta(\omega,A,\eta_2^{(k)})=0\, ,
\end{eqnarray}  
where the indices $s$, $n$, $l$, and $m$ on $\omega$ and $A$ have been omitted for notational clarity. The QNM spectra $\omega_{lm}^{(s)}$ and separation constants $A_{lm}^{(s)}$ emerge as simultaneous roots satisfying both equations.

Centering on the idea of parameterized QNMs~\cite{Cardoso:2019mqo,McManus:2019ulj,Cano:2024jkd}, both the QNM spectra and separation constants can be expanded as Taylor series around their GR corresponding values. Up to the linear order, such expansions have the following forms
\begin{eqnarray}
    \label{linear_sol}
\omega&=&\omega^{(0)}+\sum_{k=0}^{K_1}d_{1\omega}^{(k)}\eta_1^{(k)}+\sum_{k=0}^{K_2}d_{2\omega}^{(k)}\eta_2^{(k)}\, ,\nonumber\\
    A&=&A^{(0)}+\sum_{k=0}^{K_1}d_{1A}^{(k)}\eta_1^{(k)}+\sum_{k=0}^{K_2}d_{2A}^{(k)}\eta_2^{(k)}\, ,
\end{eqnarray}
where the notion $(0)$ stands for the GR case and the $d$-functions form the basis of the parametrized framework, whose explicit determination constitutes the main target of this section. In order to determine a series of $d$ above, equations~\eqref{algebratic_Eq} are expanded about $\eta=0$ via Taylor series and retain terms up to linear order. This implies that for $j=r$ or $j=\theta$, to the linear order, formally, we have
\begin{eqnarray}\label{linear_sol_expansion}
    \mathcal{L}_j(\omega^{(0)},A^{(0)},0)+\eta\frac{\partial\mathcal{L}_j}{\partial\eta}\Big(\omega^{(0)},A^{(0)},0\Big)+ \eta d_{\omega}\frac{\partial\mathcal{L}_j}{\partial\omega}\Big(\omega^{(0)},A^{(0)},0\Big)+ \eta d_{A}\frac{\partial\mathcal{L}_j}{\partial A}\Big(\omega^{(0)},A^{(0)},0\Big)+\mathcal{O}(\eta^2)=0\, .
\end{eqnarray}
Within the definitions of QNM spectra and the separation constants in GR, the first term of Eq.~\eqref{linear_sol_expansion} is vanished, i.e., $\mathcal{L}_j(\omega^{(0)},A^{(0)},0)=0$. Therefore, solving the systems (\ref{linear_sol_expansion}) yields the explicit form of the $d$-functions, namely
\begin{eqnarray}\label{d1}
   d_{1\omega}^{(k)}=\left.-\frac{\partial \mathcal{L}_r}{\partial \eta_1^{(k)}}\cdot\frac{\partial \mathcal{L}_\theta}{\partial A}\cdot\Big(\frac{\partial \mathcal{L}_r}{\partial \omega}\cdot\frac{\partial \mathcal{L}_\theta}{\partial A}-\frac{\partial \mathcal{L}_\theta}{\partial \omega}\cdot\frac{\partial \mathcal{L}_r}{\partial A}\Big)^{-1}\right|_{\mathrm{GR}}\, ,
\end{eqnarray}
\begin{eqnarray}\label{d2}   
    d_{1A}^{(k)}=\left.\frac{\partial \mathcal{L}_r}{\partial \eta_1^{(k)}}\cdot\frac{\partial \mathcal{L}_\theta}{\partial \omega}\cdot\Big(\frac{\partial \mathcal{L}_r}{\partial \omega}\cdot\frac{\partial \mathcal{L}_\theta}{\partial A}-\frac{\partial \mathcal{L}_\theta}{\partial \omega}\cdot\frac{\partial \mathcal{L}_r}{\partial A}\Big)^{-1}\right|_{\mathrm{GR}}\, ,
\end{eqnarray}
\begin{eqnarray}\label{d3}
   d_{2\omega}^{(k)}=\left.-\frac{\partial \mathcal{L}_\theta}{\partial \eta_2^{(k)}}\cdot\frac{\partial \mathcal{L}_r}{\partial A}\cdot\Big(\frac{\partial \mathcal{L}_\theta}{\partial \omega}\cdot\frac{\partial \mathcal{L}_r}{\partial A}-\frac{\partial \mathcal{L}_r}{\partial \omega}\cdot\frac{\partial \mathcal{L}_\theta}{\partial A}\Big)^{-1}\right|_{\mathrm{GR}}\, ,
\end{eqnarray}
\begin{eqnarray}\label{d4}   
    d_{2A}^{(k)}=\left.\frac{\partial \mathcal{L}_\theta}{\partial \eta_2^{(k)}}\cdot\frac{\partial \mathcal{L}_r}{\partial \omega}\cdot\Big(\frac{\partial \mathcal{L}_\theta}{\partial \omega}\cdot\frac{\partial \mathcal{L}_r}{\partial A}-\frac{\partial \mathcal{L}_r}{\partial \omega}\cdot\frac{\partial \mathcal{L}_\theta}{\partial A}\Big)^{-1}\right|_{\mathrm{GR}}\, ,
\end{eqnarray}
where GR means that $\omega=\omega^{(0)}$, $A=A^{(0)}$, and all $\eta_{1,2}^{(k)}=0$.

\begin{figure}
    \hspace{-2cm}
    \includegraphics[width=0.81\textheight]{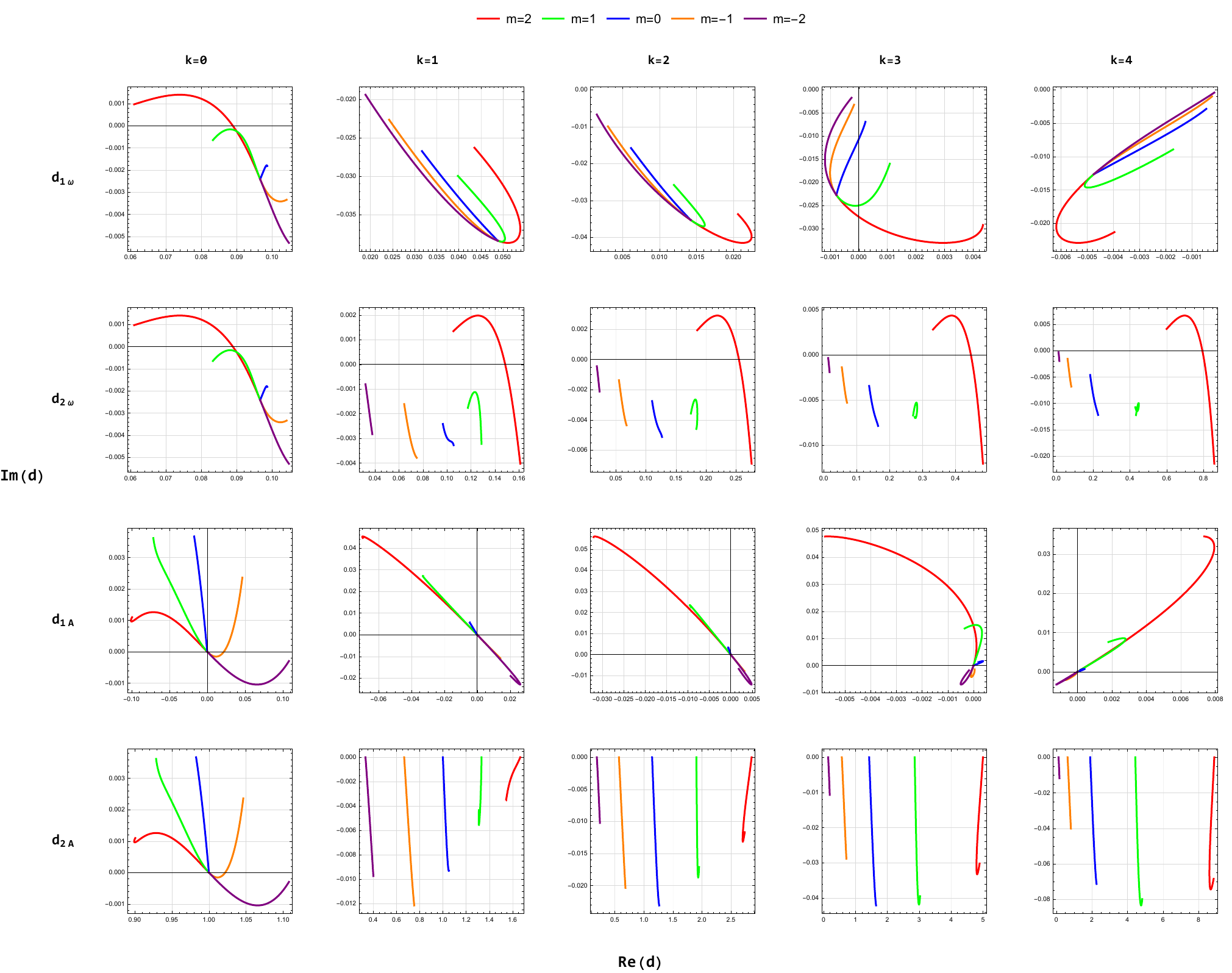}
    \caption{The figure represents the $d$-functions for tensor modes as functions of the parameter $a$ in the range $[0,0.45]$, presented for $l=2$. Different colors represent different values of $m$. Note that $d^{(0)}_{1\omega}$ and $d^{(0)}_{2\omega}$ yield identical results due to the derivative structure in their defining equations when $k=0$.}
\label{fig_d}
\end{figure}

As a typical example, for for $l=2$ tensor modes $(s=-2)$, the numerical results for the $d$-functions in Eqs. \eqref{d1}-\eqref{d4} are shown in Fig.~\ref{fig_d} as functions of the angular momentum parameter $a$ over the range $[0,0.45]$. This range is chosen because as $a$ increases, the event horizon $r_+$ approaches the Cauchy horizon $r_-$, causing the solution ansatz \eqref{R_Structure} to become increasingly inaccurate near the extremal limit $a=0.5$. In terms of numerical details, numerical derivatives are computed using a six-point finite difference scheme, and convergence is ensured by adjusting the truncation order $n$ until relative changes fell below $10^{-7}$. While $K_1$ and $K_2$ are generally theory-dependent, we fix $K_1=K_2=4$ for present analysis (There are $25$ panels in Fig. \ref{fig_d}.). Notably, $d_{1\omega}^{(0)}$ and $d_{2\omega}^{(0)}$ yield identical values due to the derivative structure in Eqs. \eqref{d1} and \eqref{d3} acting solely on $\mathbf{\Theta}^{(0)}_n$ and $\mathbf{R}^{(0)}_n$ when $k=0$, resulting in the identity $\partial\mathcal{L}_r/\partial\eta_1^{(0)}\cdot\partial\mathcal{L}_\theta/\partial A = -\partial\mathcal{L}_\theta/\partial\eta_2^{(0)}\cdot\partial\mathcal{L}_r/\partial A$. Additionally, all $d_{1\omega}^{(k)}$ and $d_{1A}^{(k)}$ with different azimuthal numbers $m$ coincide at $a=0$, since the radial recurrence relation \eqref{recurrence_radial} depends exclusively on the product $am$. This degeneracy is absent for the angular sector, except for $d_{2\omega}^{(0)}$ and $d_{2A}^{(0)}$, where $\partial\mathcal{L}_\theta/\partial\eta_{2}^{(0)}|_{\mathrm{GR}}$ becomes constant as $a\to 0$ and $A\to0$ when $-2\leq m \leq2$. The properties and functional forms of the $d$-functions are intrinsic to the modified TE \eqref{TE_with_perturbation} under the ansatz \eqref{ansatz_separation}. The deviation parameters can be directly determined from QNM observational data through Eq. \eqref{linear_sol}, and we will leave the constraint on the parameter $\eta_1^{(k)}$ and $\eta_2^{(k)}$ by utilizing existing observational data in the future~\cite{Chen:2025wfi}.


\section{QNM spectra within two-dimensional pseudo-spectral method}
\label{Sec: QNM_two_dimensioncal}
In Sec. \ref{Sec: Parameterized_framework}, we have established a parameterized framework for QNMs for the modified TE~\eqref{TE_with_perturbation} or ~\eqref{BL_TE_with_perturbation}, but it is clearly only a linear order approximation. A natural question to ask is, how effective is this approximation? This prompts us to accurately derive the corresponding QNM spectra. This section presents direct computation of the QNM spectra for the modified TE \eqref{BL_TE_with_perturbation} using the two-dimensional pseudo-spectral method~\cite{Jaramillo:2020tuu,Xiong:2024urw,Cai:2025irl,Assaad:2025nbv}, providing independent cross-validation and deriving corresponding parameter constraints to verify consistency at first order. 

For the modified TE within the B-L coordinates, since all the constant-$t$ surfaces intersect the spatial infinity $i^0$ and the bifurcation sphere $\mathcal{B}$ simultaneously, the QNM eigenfunctions exhibit divergent behavior. This necessitates careful imposition of QNM boundary conditions through prior separation of the divergent components. In contrast, the hyperboloidal framework employs constant-$\tau$ surfaces that extend smoothly between future null infinity $\mathscr{I}^+$ and the future event horizon $\mathscr{H}^+$, ensuring the regularity of QNMs~\cite{Ansorg:2016ztf,PanossoMacedo:2018hab,PanossoMacedo:2019npm,Jaramillo:2020tuu,PanossoMacedo:2023qzp,PanossoMacedo:2024nkw,Zhou:2025xta,Cai:2025irl,Assaad:2025nbv,Ripley:2022ypi}. Note that the hyperboloidal framework has been achieved in many black holes~\cite{Jaramillo:2020tuu,Destounis:2021lum,Cao:2024oud,Wu:2024ldo,Cao:2024sot,Yang:2025hqk,PanossoMacedo:2019npm}. Note that when the master equation has a correction term of the velocity, the corresponding hyperbolidal framework can still work~\cite{Cao:2025qws}. See also some comments on the hyperboloidal framework for black hole QNMs~\cite{Shen:2025nbq}. This hyperboloidal approach consequently enables both independent numerical computation of QNMs and detailed investigations of the spectrum instability or the pseudospectrum. One can refer to the study of the pseudospectrum within the hyperboloidal framework in~\cite{Jaramillo:2020tuu,Cai:2025irl,Cao:2024oud,Chen:2024mon,Cao:2024sot,Cao:2025qws,Destounis:2023nmb,Cownden:2023dam,Boyanov:2023qqf} and therein.

For the Kerr black hole, the hyperboloidal framework is completed by~\cite{PanossoMacedo:2019npm}. Although the hyperboloidal framework there was originally constructed for the original TE, it can still be employed for our modified TE (\ref{TE_with_perturbation}). This is due to the fact that the modifications (\ref{eta_2}) and (\ref{eta_1}) do not alter the asymptotic characteristics of the QNM functions, including both their radial and angular parts. The complete mappings from Boyer-Lindquist to the hyperboloidal coordinates are given by
\begin{eqnarray}\label{coordinate_transformation}
    t=\lambda\Big[\tau-h(\sigma,\theta)\Big]-r_{\star}(r(\sigma))\, ,\quad r(\sigma)=\lambda\frac{\rho(\sigma)}{\sigma}\, ,\quad \varphi=\phi-k(r(\sigma))\, ,
\end{eqnarray}
in which the tortoise function $r_{\star}(r)$ and the phase function $k(r)$ are defined by
\begin{eqnarray}
    \frac{\mathrm{d}r_{\star}}{\mathrm{d}r}=\frac{\Sigma_0}{\Delta}\, ,\quad \frac{\mathrm{d}k}{\mathrm{d}r}=\frac{a}{\Delta}\, .
\end{eqnarray}
Here, $h(\sigma,\theta)$ is referred to as the height function. Together with the radial function $\rho(\sigma)$, they characterize the degrees of freedom of the gauge, and their explicit forms depend on the specific gauge choice. In this study, choosing $\lambda=r_{+}$, we adopt the minimal gauge with radial function fixing~\cite{PanossoMacedo:2019npm}, i.e., 
\begin{eqnarray}\label{minimal_gauge_radial_function_fixing}
    \rho(\sigma)=1\, ,\quad\beta(\sigma)=1\, ,\quad h(\sigma, \theta)=-\frac{2}{\sigma}+4\mu \ln\sigma\, ,\quad \mu=\frac{1+\alpha^2}{2}\, ,
\end{eqnarray}
where the dimensionless mass $\mu$ and dimensionless spin $\alpha$ are
\begin{eqnarray}
    \mu=M/\lambda\, ,\quad\alpha=a/\lambda\, .
\end{eqnarray}

To regularize the essential singularities at future null infinity $\sigma=0$ and the black hole horizon $\sigma=\sigma_{+}$, while ensuring regularity at the polar boundaries $\sin\theta=0$, the master function $\Psi^{(s)}$ of Eq. (\ref{BL_TE_with_perturbation}) is written as~\cite{PanossoMacedo:2019npm}
\begin{eqnarray}
     \Psi^{(s)}(\tau,\sigma,\theta,\phi)=\Omega\Big[\Delta(\sigma)\Big]^{-s}\sum_{m=-\infty}^{+\infty}\cos^{\delta_1^s}(\theta/2)\sin^{\delta_2^s}(\theta/2)V^{(s)}_m(\tau,\sigma,\theta)\mathrm{e}^{\mathrm{i}m\phi}\, ,
\end{eqnarray}
in which the exponents are $\delta_1^s=|m-s|$ and $\delta_2^s=|m+s|$, with $\Delta(\sigma)=\Delta(r(\sigma))$. The conformal factor is given by $\Omega=\sigma/\lambda$, while the azimuthal dependence takes the form $\mathrm{e}^{\mathrm{i}m\phi}$ . By introducing the substitution $x=\cos\theta$ and and performing a first-order time reduction through $W_m=\partial_\tau V_m$, one can recast the Eq. (\ref{BL_TE_with_perturbation}) as a set of partial differential equations involving first-order temporal derivatives and second-order spatial derivatives, i.e.,
\begin{eqnarray}\label{dynamics_eq}
	\partial_\tau \begin{bmatrix}
		V_m(\tau,\sigma,x)\\
		W_m(\tau,\sigma,x)
	\end{bmatrix}=\mathrm{i}L(\sigma,x)\begin{bmatrix}
		V_m(\tau,\sigma,x)\\
		W_m(\tau,\sigma,x)
	\end{bmatrix}\, ,\quad \text{with} \quad  L(\sigma,x)=\frac{1}{\mathrm{i}}
	\begin{bmatrix}
		0 & 1\\
		L_1(\sigma,x) & L_2(\sigma,x)
	\end{bmatrix}\, ,
\end{eqnarray}
where the operator $L(\sigma,x)$ can be looked as the time generator of the linear dynamics and encode the boundary conditions. The explicit forms of $L_1(\sigma,x)$ and $L_2(\sigma,x)$ are provided in Appendix \ref{app:2}. Transforming the system to the frequency domain via separating the time part $\mathrm{e}^{\mathrm{i}\omega \tau}$ yields the two-dimensional eigenvalue problem
\begin{eqnarray}\label{QNM_eigenvalue_problem}
	L(\sigma,x)\begin{bmatrix}
		\mathbb{V}_m(\omega,\sigma,x)\\
		\mathbb{W}_m(\omega,\sigma,x)
	\end{bmatrix}=\omega
	\begin{bmatrix}
		\mathbb{V}_m(\omega,\sigma,x)\\
		\mathbb{W}_m(\omega,\sigma,x)
	\end{bmatrix}\, ,
\end{eqnarray}
Within the help of hyperboloidal framework, the eigenfunctions of the spectral problem are regular in the region $(\sigma,x)\in[0,1]\times[-1,1]$.  We will use a two-dimensional pseudo-spectral method to solve the eigenvalue problem in a tensor product grid. The details of such a method can be found in~\cite{Jansen:2017oag,Jaramillo:2020tuu,Xiong:2024urw,Cai:2025irl,doi:10.1137/1.9780898719598,Miguel:2023rzp,Assaad:2025nbv}. For each direction grid, the Chebyshev-Lobatto grid is used, where the Chebyshev-Lobatto grid is given by
\begin{eqnarray}
    \sigma_i&=&\frac{1}{2}\Big[1+\cos\Big(\frac{i\pi}{N_\sigma}\Big)\Big]\, ,\quad i=0,1,\cdots,N_\sigma-1,N_\sigma\ ,\nonumber\\
    x_j&=&\cos\Big(\frac{j\pi}{N_x}\Big)\, ,\quad j=0,1,\cdots,N_x-1, N_x\, ,
\end{eqnarray}
and $N_\sigma$ is the resolution for the $\sigma$ direction, $N_x$ is the resolution for the $x$ direction. Therefore, the dimension of the finite rank approximation $\mathbf{L}$, a square matrix, for the original operator $L$ is $2(N_\sigma+1)(N_x+1)\times2(N_\sigma+1)(N_x+1)$.

\begin{figure}
    \hspace{-2cm}
    \includegraphics[width=0.81\textheight]{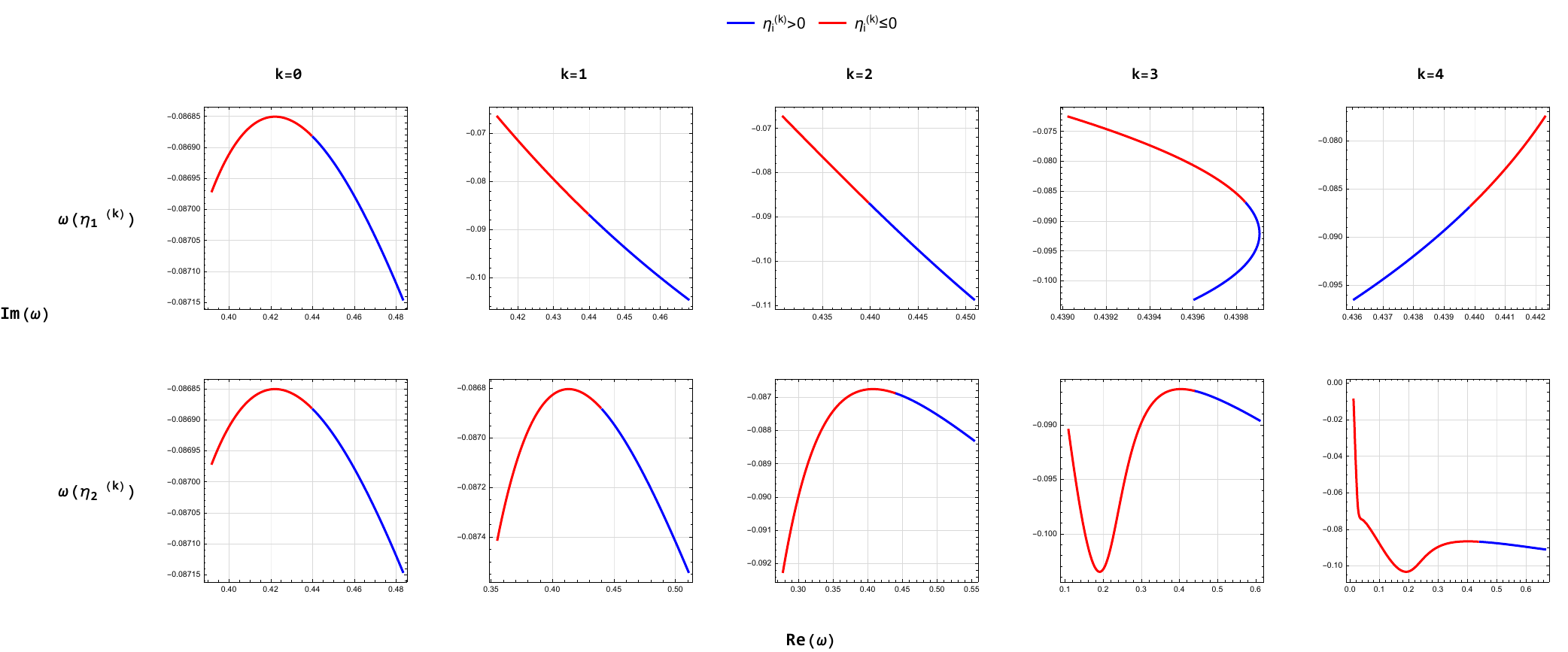}
    \caption{The figure displays the migrations of QNM spectra as functions of the deviation parameters for the fundamental modes with $l=2$,$m=2$ and angular momentum parameter $a=0.2$. The first row illustrates the dependence on $\eta_{1}^{(k)}$, while the second row shows the corresponding variations with $\eta_{2}^{(k)}$.  In all panels, the blue curve segment corresponds to the trajectory for $\eta_{i}^{(k)}>0$, and the red curve segment to $\eta_{i}^{(k)}<0$. The migration trajectory for $\omega(\eta^{(1)}_1)$ coincides with that for 
$\omega(\eta_2^{(1)})$.}
\label{fig2}
\end{figure}

\begin{figure}
    \hspace{-2cm}
    \includegraphics[width=0.75\textheight]{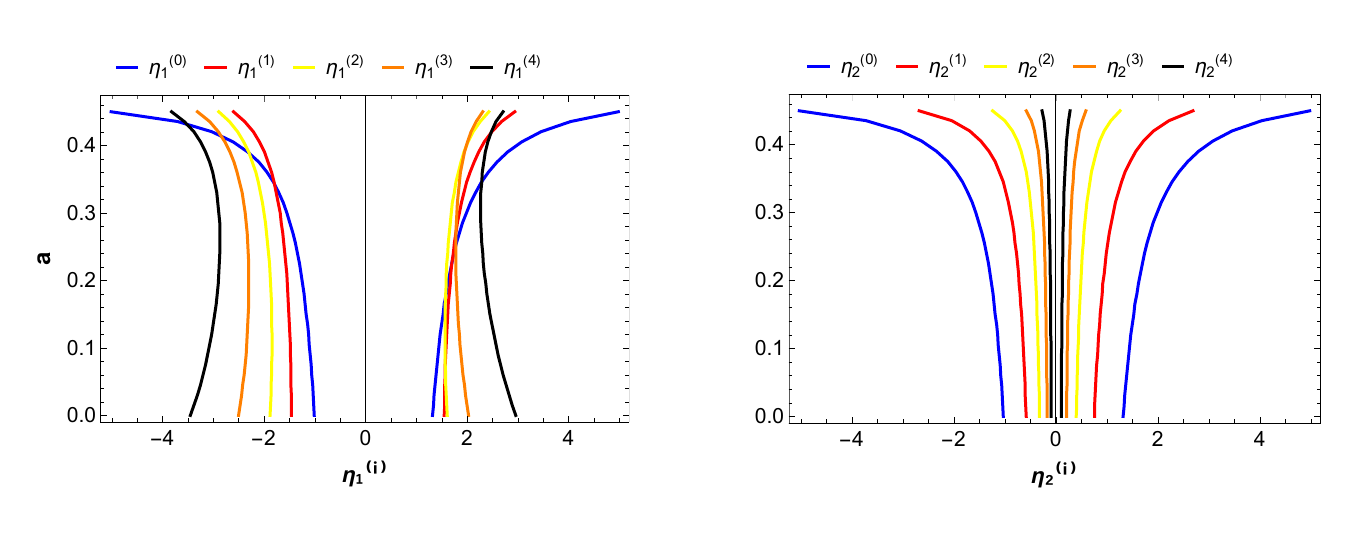}
    \caption{Constraints on the deviation parameters obtained by requiring agreement between the continued fraction and pseudo-spectral methods within $1\%$, shown as functions of the black hole spin parameter 
    $a\in[0,0.45]$, where blue lines refer to $\eta_i^{(0)}$, red lines refer to $\eta_i^{(1)}$, yellow lines refer to $\eta_i^{(2)}$, orange lines refer to $\eta_i^{(3)}$, and black lines refer to $\eta_i^{(4)}$.}
    \label{fig3}
\end{figure}
To determine the final QNM spectra, we systematically vary the grid resolutions $N_x$ and $N_\sigma$ until numerical convergence is achieved. Fig. \ref{fig2} shows the migrations of the fundamental QNM spectra with $l=2$, $m=2$ and $\text{Re}(\omega)>0$ as the deviation parameters are varied. Note that each panels are derived by only varying one $\eta_1^{(k)}$ or $\eta_2^{(k)}$. The parameters $\eta_i^{(k)}$ are restricted to the range $(-1,1)$ to maintain the validity of the linear approximation \eqref{linear_sol}. Notably,  $\omega(\eta_1^{(0)})$ and $\omega(\eta_2^{(0)})$ exhibit identical trajectories, consistent with the results obtained via the continued fraction method where the corresponding $d$-functions coincide. 

To validate the QNM sepctra part of the parameterized framework~\eqref{linear_sol}, we derive constraints on the deviation parameters, where these bounds are determined by requiring agreement within $1\%$ between the continued fraction method and the two-dimensional pseudospectral method. The relative error is calculated via $\lvert\omega_1-\omega_2\rvert/\lvert\omega_1\rvert$, where $\omega_1$ and $\omega_2$ denote the results in pseudo-spectral method and the linear QNM framework~\eqref{linear_sol}  from the continued fraction method, respectively. The compared results presented in Fig.~\ref{fig3}, where the color lines correspond to threshold $1\%$. From this figure, it can be found that most parameters exhibit bounds satisfy $|\eta|\ge1$, confirming both the validity of the linear approximation in Eq.~\eqref{linear_sol} and the accuracy of the continued fraction method. The exceptions are $\eta_2^{(3)}$ and $\eta_2^{(4)}$, where rapid variations in their trajectories lead to tighter constraints. These variations are primarily driven by the large real parts of $d_{2\omega}^{(3)}$ and $d_{2\omega}^{(4)}$. Notably, the parameter bounds broaden with increasing $a$, contrasting with the behavior reported in Ref.~\cite{Cano:2024jkd}.

\section{Discussion and Conclutions}
\label{sec:Conlusion}
In this work, we have established a parameterized framework for the modified Teukolsky equation, introducing a separable deviation term to capture agnostic modifications from GR. This approach leads to decoupled radial and angular equations, where the modifications in both sectors are parameterized by independent sets of deviation parameters. We successfully solve the obtained system using the semi-analytical continued fraction method, deriving the linear corrections to the QNM spectra and the separation constants to construct the parameterized framework [see Eq. \eqref{linear_sol}]. This extends the parameterized QNM formalism from spherical symmetry to axisymmetric spacetimes, while self-consistently treating deviations in both the radial and angular sectors.

The robustness of our results was secured through independent cross-validation with the two-dimensional pseudo-spectral method. This numerical approach directly solves the eigenvalue problem, natively handling the full two-dimensional nature of the perturbations without relying on separability. The excellent agreement between the continued fraction and pseudo-spectral results confirms the self-consistency of our framework at first order in the deviation parameters, where the consistent results are shown in Fig. \ref{fig3}. Furthermore, this comparison allows us to derive constraints on the parameters, validating the linear approximation across a physically relevant range and establishing the pseudo-spectral method as a powerful tool for future studies of non-separable modifications.

In conclusion, this work provides a foundational and theory-agnostic framework for interpreting gravitational wave ringdown signals in the context of beyond-GR physics. The explicit form of the linear corrections enables efficient data analysis to search for or constrain potential deviations. Prospectively, this formalism serves as a versatile platform for future exploration. A crucial next step is to systematically map specific modified theories of gravity and environmental effects onto our deviation parameters, thereby identifying which new physics scenarios can be effectively probed with this parameterization. By bridging the gap between theoretical models and observational data, our framework may enhance the toolkit for testing GR and probing new physics in the strong-field regime with next-generation gravitational-wave detectors.

\section*{Acknowledgement}
We are grateful to Li-Ming Cao, Yu-Sen Zhou, and Ming-Fei Ji for helpful discussions. This work is supported by the National Natural Science Foundation of China. Specifically, Zhe Yu is supported by Grant No. 12447176, and Liang-Bi Wu is supported by Grant No. 12505067.

\appendix
\section{Derivation of the recurrence relations}\label{app:1}
This section presents the detailed derivation of recurrence relations for the radial and angular equations \eqref{equation_R} and \eqref{equation_S} within the continued fraction method. Introducing the variable $y=1+x$ and substituting the solution ansatz into Eq. \eqref{equation_S}, then the equation can be rewritten as 
\begin{eqnarray}
    &&\Big[y(2-y)\frac{\mathrm{d}^2}{\mathrm{d}y^2}-2(y-1)\frac{\mathrm{d}}{\mathrm{d}y}+a^2\omega^2(y-1)^2-2a\omega s(y-1)+s+A_{lm}^{(s)}(a\omega)\nonumber\\
    &&-\frac{[m+s(y-1)]^2}{y(2-y)}-\sum_{k=0}^{K_2}\eta_2^{(k)}y^k\Big]\Big[y^{k_1}(2-y)^{k_2}e^{a\omega(y-1)}\sum_{n=0}^{\infty}S_ny^n\Big]=0\, .
\end{eqnarray}
For the general case where $\eta_2^{(K_2)}\neq0$, eliminating some overall factor leads to the following form
\begin{eqnarray}
    &&\Big(U_{00}+U_{01}y+U_{02}y^2+U_{03}y^3+\cdots+U_{0K_2}y^{K_2}\Big)\sum_{n=0}^{\infty}S_ny^n+\Big(U_{10}+U_{11}y+U_{12}y^2\Big)\sum_{n=0}^{\infty}(n+1)S_{n+1}y^n\nonumber\\
    &&+\Big(U_{21}y+U_{22}y^2\Big)\sum_{n=0}^{\infty}(n+2)(n+1)S_{n+2}y^n=0\, ,
\end{eqnarray}
where the explicit forms of the various $U$-functions are omitted for brevity. By reindexing the summation terms and extracting the coefficient of $y^n$, the recurrence relation for $S_n$ can be derived as
\begin{eqnarray}    &&\epsilon[n]\Big[(n+1)U_{10}+n(n+1)U_{21}\Big]S_{n+1}+\epsilon[n]\Big[U_{00}+nU_{11}+n(n-1)U_{22}\Big]S_n+\epsilon[n-1]\Big[U_{01}+(n-1)U_{12}\Big]S_{n-1}\nonumber\\
    &&+\epsilon[n-2]U_{02}S_{n-2}+\epsilon[n-3]U_{03}S_{n-3}+\epsilon[n-4]U_{04}S_{n-4}+\cdots+\epsilon[n-K_2]U_{0K_2}S_{n-K_2}=0\, ,\quad n\ge0\, ,
\end{eqnarray}
by systematically rearranging the above terms, we obtain the angular recurrence relations~\eqref{recurrence_angular}.

For the radial sector, the solution method follows the same approach as in the angular case, though with greater complexity. To simplify the analysis, we introduce a new master field
\begin{eqnarray}
    Y(r)=\Delta^{s/2}(r^2+a^2)^{1/2}R(r)\, ,
\end{eqnarray}
which yields a more tractable form of the radial equation. Multiplying both sides of the equation by the factor
\begin{eqnarray}
    \frac{\Delta^{1+s/2}}{(r^2+a^2)^{3/2}}\, ,
\end{eqnarray}
we obtain the following differential equation for $Y(r)$
\begin{eqnarray}\label{equation_Y}
    &&\frac{\Delta^2}{(a^2+r^2)^2}\frac{\mathrm{d}^2Y(r)}{\mathrm{d}r^2}+\frac{\Delta[(a^2+r^2) \Delta^{\prime}-2 r \Delta]}{(a^2+r^2)^{3}}\frac{\mathrm{d}Y(r)}{\mathrm{d}r}+\Bigg\{\frac{1}{4(a^2+r^2)^{4}}\Big[-4(a^2-2r^2) \Delta^2-s^2(a^2+r^2)^2 (\Delta^{\prime})^2\nonumber\\
    &&-2(a^2+r^2)\Delta\Big(s(a^2+r^2) \Delta^{\prime\prime}+2 r \Delta^{\prime}\Big)\Big]\Bigg\}Y(r)+\frac{\Delta[V(r)-\eta_1(r)]}{(a^2+r^2)^2}Y(r)=0\, .
\end{eqnarray}
To analyze the solution structure, we begin by examining the asymptotic behavior of the solution. In the limit $r\to\infty$, where higher-order terms can be neglected, the equation reduces to
\begin{eqnarray}
    \frac{\mathrm{d}^2Y}{\mathrm{d}r_{\star}^2}+\Big(\omega^2+\frac{2\mathrm{i}s\omega}{r}\Big)Y=0\, ,
\end{eqnarray}
where the tortoise coordinate $r_{\star}$ is introduced to obtain this standard form. According to the boundary conditions for QNMs, the function $Y$ can be solved as
\begin{eqnarray}
    Y\sim r^{p-s}e^{q r}\, ,\quad p\equiv2\mathrm{i}M\omega\, ,\quad q\equiv \mathrm{i}\omega\, .
\end{eqnarray}
Correspondingly, the asymptotic behavior of the radial function $R$ is given by
\begin{eqnarray}
    R\sim r^{p-2s-1}e^{qr}\, ,\quad r\to\infty\, .
\end{eqnarray}

As $r\to r_{+}$, the tortoise coordinate $r_{\star}$ defined in Eq. \eqref{tortoise_coordinate} behaves asymptotically as
\begin{eqnarray}
    r_{\star}\to\frac{r_{+}}{\beta}\ln(r-r_{+})\, .
\end{eqnarray}
In this limit, the radial equation reduces to
\begin{eqnarray}
    \frac{\mathrm{d}^2Y}{\mathrm{d}r_{\star}^2}+\Big(\frac{K_{+}}{2Mr_{+}}-\frac{\mathrm{i} s\beta}{2r_{+}}\Big)^2Y(r)=0\, ,\quad K_{+}\equiv 2\omega Mr_{+}-am\, .
\end{eqnarray}
Imposing the purely ingoing wave condition at the horizon, we obtain the solution
\begin{eqnarray}
    Y&\sim&\exp{\Bigg[-\mathrm{i}r_{\star}\Big(\frac{K_{+}}{2Mr_{+}}-\frac{\mathrm{i} s\beta}{2r_{+}}\Big)\Bigg]}\sim(r-r_{+})^{-\mathrm{i}\sigma-s/2}\, ,\quad \sigma\equiv\frac{K_{+}}{2\beta M}\, .
\end{eqnarray}
Therefore, the asymptotic behavior of $R$ is
\begin{eqnarray}
    R\sim (r-r_{+})^{-\mathrm{i}\sigma-s}\, ,\quad r\to r_{+}\, .
\end{eqnarray}
We introduce a new variable
\begin{eqnarray}
f=\frac{r-r_{+}}{r-r_{-}}\, ,\quad \text{and} \quad   r=\frac{M(1+\beta-f+f\beta)}{1-f}\, ,
\end{eqnarray}
and substitute this transformation into the radial equation~\eqref{equation_R}. The resulting equation takes the following form
\begin{eqnarray}
    &&\Bigg\{(1-f)^2f\frac{\mathrm{d}^2}{\mathrm{d}f^2}+(1-f)\Big[s+1+(s-1)f\Big]\frac{\mathrm{d}}{\mathrm{d}f}+V(f)-\sum_{k=0}^{K_1}\eta_1^{(k)}\sum_{j=0}^k\binom{k}{j}(-1)^jf^j\Bigg\}\nonumber\\
    &&\Bigg\{f^{-\mathrm{i}\sigma-s}\Big(\frac{2M\beta}{1-f}\Big)^{p-2s-1}\exp\Big[q\Big(\frac{M(1+\beta-f+f\beta)}{1-f}\Big)\Big]\sum_{n=0}^{\infty}R_nf^n\Bigg\}=0\, .
\end{eqnarray}
For the general case with nonvanishing $\eta_1^{(K_1)}$, removal of an overall factor yields
\begin{eqnarray}
    &&\Big(U_{00}+U_{01}f+U_{02}f^2+U_{03}f^3+\cdots+U_{0K_1}f^{K_1}\Big)\sum_{n=0}^{\infty}R_nf^n+\Big(U_{10}+U_{11}f+U_{12}f^2\Big)\sum_{n=0}^{\infty}(n+1)R_{n+1}f^n\nonumber\\
    &&+\Big(U_{21}f+U_{22}f^2+U_{23}f^3\Big)\sum_{n=0}^{\infty}(n+2)(n+1)R_{n+2}f^n=0\, .
\end{eqnarray}
Similarly, the details of the $U$-functions above are omitted. Changing the indicators for each sums and considering the coefficient of $f^n$, we obtain the recurrence relation of $R_n$ as follow
\begin{eqnarray}
    &&\epsilon[n]\Big[(n+1)U_{10}+n(n+1)U_{21}\Big]R_{n+1}+\epsilon[n]\Big[U_{00}+nU_{11}+n(n-1)U_{22}\Big]R_n\nonumber\\
    &&+\epsilon[n-1]\Big[U_{01}+(n-1)U_{12}+(n-1)(n-2)U_{23}\Big]R_{n-1}+\epsilon[n-2]U_{02}R_{n-2}\nonumber\\
    &&+\epsilon[n-3]U_{03}R_{n-3}+\cdots+\epsilon[n-K_1]U_{0K_1}R_{n-K_1}=0\, ,\quad n\ge0\, .\nonumber
\end{eqnarray}
After rearranging the above terms, we will obtain the radial recurrence relations~\eqref{recurrence_radial}.

\section{The expressions of the operator $L_1$ and $L_2$}\label{app:2}
In this appendix, for the radial function fixing guage~\cite{PanossoMacedo:2019npm} given by Eqs. (\ref{minimal_gauge_radial_function_fixing}), we will give the explicit expression of the operator $L$ for the modified TE (\ref{BL_TE_with_perturbation}). The expressions of the operators $L_1(\sigma,x)$ and $L_2(\sigma,x)$ are written as
\begin{eqnarray}
	L_1(\sigma,x)={}^{[x]}L_1^{2}(\sigma,x)\frac{\partial^2}{\partial x^2}+{}^{[\sigma]}L_1^{2}(\sigma,x)\frac{\partial^2}{\partial \sigma^2}+{}^{[x]}L_1^{1}(\sigma,x)\frac{\partial}{\partial x}+{}^{[\sigma]}L_1^{1}(\sigma,x)\frac{\partial}{\partial \sigma}+L_1^0(\sigma,x)\, ,
\end{eqnarray}
\begin{eqnarray}
	L_2(\sigma,x)={}^{[\sigma]}L_2^{1}(\sigma,x)\frac{\partial}{\partial \sigma}+L_2^0(\sigma,x)\, ,
\end{eqnarray}
where seven functions ${}^{[x]}L_1^{2}(\sigma,x)$, ${}^{[\sigma]}L_1^{2}(\sigma,x)$, ${}^{[x]}L_1^{1}(\sigma,x)$, ${}^{[\sigma]}L_1^{1}(\sigma,x)$, $L_1^0(\sigma,x)$, ${}^{[\sigma]}L_2^{1}(\sigma,x)$, and $L_2^0(\sigma,x)$ associated with $(\sigma,x)$ are
\begin{eqnarray}
	{}^{[x]}L_1^{2}(\sigma,x)=\frac{1-x^2}{4(\alpha ^2+1)[\alpha ^2 (1-\sigma )+1] [(\alpha^2+1) \sigma +1]-\alpha ^2(1-x^2)}\, ,
\end{eqnarray}
\begin{eqnarray}
	{}^{[\sigma]}L_1^{2}(\sigma,x)=\frac{(1-\sigma ) \sigma ^2(1-\alpha ^2 \sigma)}{4(\alpha ^2+1)[\alpha^2 (1-\sigma)+1][(\alpha ^2+1) \sigma +1]-\alpha ^2(1-x^2)}\, ,
\end{eqnarray}
\begin{eqnarray}
	{}^{[x]}L_1^{1}(\sigma,x)=\frac{\delta^s_1 (1-x)-\delta^s_2(x+1)-2 x}{4(\alpha ^2+1)[\alpha ^2 (1-\sigma )+1][(\alpha ^2+1) \sigma +1]-\alpha ^2(1-x^2)}\, ,
\end{eqnarray}
\begin{eqnarray}
	{}^{[\sigma]}L_1^{1}(\sigma,x)=\frac{\sigma\Big(2 (s+1)-\sigma[-4 \alpha ^2 \sigma +2 \mathrm{i} \alpha  m+(\alpha ^2+1) (s+3)]\Big)}{4(\alpha ^2+1)[\alpha ^2 (1-\sigma)+1][(\alpha^2+1) \sigma +1]-\alpha ^2(1-x^2)}\, ,
\end{eqnarray}
\begin{eqnarray}\label{L_1_0}
	L_1^0(\sigma,x)=\frac{-\sigma [\alpha^2(1-2\sigma )+1]-2\mathrm{i}\alpha m\sigma-(\alpha ^2+1) s \sigma +\Big(\frac{\delta^s_1+\delta^s_2}{2}+s+1\Big) \Big(-\frac{\delta^s_1+\delta^s_2}{2}+s\Big)-\delta L_1^0(\sigma,x)}{4(\alpha ^2+1)[\alpha ^2 (1-\sigma )+1][(\alpha ^2+1) \sigma +1]-\alpha ^2(1-x^2)}\, ,
\end{eqnarray}
\begin{eqnarray}
	{}^{[\sigma]}L_2^{1}(\sigma,x)=\frac{2\Big(\alpha ^2\sigma^2-2(\alpha ^2+1) \sigma ^2[\alpha^2 (1-\sigma)+1]+1\Big)}{4(\alpha ^2+1)[\alpha ^2 (1-\sigma )+1] [(\alpha ^2+1) \sigma +1]-\alpha ^2(1-x^2)}\, ,
\end{eqnarray}
\begin{eqnarray}
	L_2^0(\sigma,x)=\frac{2 \Big(\sigma[\alpha ^4 (2-3 \sigma )-3 \alpha ^2 (\sigma -1)+2]+\mathrm{i} m(2 \alpha ^3 \sigma +2 \alpha  \sigma +\alpha)+s[(\alpha ^2+1)(\alpha^2\sigma +\sigma -1)+\mathrm{i}\alpha x]\Big)}{4(\alpha ^2+1)[\alpha ^2 (1-\sigma )+1] [(\alpha ^2+1) \sigma +1]-\alpha ^2(1-x^2)}\, .
\end{eqnarray}
In Eq. (\ref{L_1_0}), the term $\delta L_1^0(\sigma,x)$ is 
\begin{eqnarray}
    \delta L_1^0(\sigma,x)=\Big(\frac{1}{\sigma^2}+\alpha^2x^2\Big)\eta(\sigma,x)=\sum _{k=0}^{K_2} \eta_2^{(k)} (1+x)^k+\sum_{k=0}^{K_1} \eta_1^{(k)} \Big[\frac{(\alpha
   ^2-1) \sigma }{\alpha ^2 \sigma
   -1}\Big]^k\, .
\end{eqnarray}

\bibliography{reference}

\begin{thebibliography}{70}%
\makeatletter
\providecommand \@ifxundefined [1]{%
 \@ifx{#1\undefined}
}%
\providecommand \@ifnum [1]{%
 \ifnum #1\expandafter \@firstoftwo
 \else \expandafter \@secondoftwo
 \fi
}%
\providecommand \@ifx [1]{%
 \ifx #1\expandafter \@firstoftwo
 \else \expandafter \@secondoftwo
 \fi
}%
\providecommand \natexlab [1]{#1}%
\providecommand \enquote  [1]{``#1''}%
\providecommand \bibnamefont  [1]{#1}%
\providecommand \bibfnamefont [1]{#1}%
\providecommand \citenamefont [1]{#1}%
\providecommand \href@noop [0]{\@secondoftwo}%
\providecommand \href [0]{\begingroup \@sanitize@url \@href}%
\providecommand \@href[1]{\@@startlink{#1}\@@href}%
\providecommand \@@href[1]{\endgroup#1\@@endlink}%
\providecommand \@sanitize@url [0]{\catcode `\\12\catcode `\$12\catcode
  `\&12\catcode `\#12\catcode `\^12\catcode `\_12\catcode `\%12\relax}%
\providecommand \@@startlink[1]{}%
\providecommand \@@endlink[0]{}%
\providecommand \url  [0]{\begingroup\@sanitize@url \@url }%
\providecommand \@url [1]{\endgroup\@href {#1}{\urlprefix }}%
\providecommand \urlprefix  [0]{URL }%
\providecommand \Eprint [0]{\href }%
\providecommand \doibase [0]{http://dx.doi.org/}%
\providecommand \selectlanguage [0]{\@gobble}%
\providecommand \bibinfo  [0]{\@secondoftwo}%
\providecommand \bibfield  [0]{\@secondoftwo}%
\providecommand \translation [1]{[#1]}%
\providecommand \BibitemOpen [0]{}%
\providecommand \bibitemStop [0]{}%
\providecommand \bibitemNoStop [0]{.\EOS\space}%
\providecommand \EOS [0]{\spacefactor3000\relax}%
\providecommand \BibitemShut  [1]{\csname bibitem#1\endcsname}%
\let\auto@bib@innerbib\@empty
\bibitem [{\citenamefont {Abbott}\ \emph {et~al.}(2023)\citenamefont {Abbott}
  \emph {et~al.}}]{KAGRA:2021vkt}%
  \BibitemOpen
  \bibfield  {author} {\bibinfo {author} {\bibfnamefont {R.}~\bibnamefont
  {Abbott}} \emph {et~al.} (\bibinfo {collaboration} {KAGRA, VIRGO, LIGO
  Scientific}),\ }\href {\doibase 10.1103/PhysRevX.13.041039} {\bibfield
  {journal} {\bibinfo  {journal} {Phys. Rev. X}\ }\textbf {\bibinfo {volume}
  {13}},\ \bibinfo {pages} {041039} (\bibinfo {year} {2023})},\ \Eprint
  {http://arxiv.org/abs/2111.03606} {arXiv:2111.03606 [gr-qc]} \BibitemShut
  {NoStop}%
\bibitem [{\citenamefont {Nollert}(1999)}]{Nollert:1999ji}%
  \BibitemOpen
  \bibfield  {author} {\bibinfo {author} {\bibfnamefont {H.-P.}\ \bibnamefont
  {Nollert}},\ }\href {\doibase 10.1088/0264-9381/16/12/201} {\bibfield
  {journal} {\bibinfo  {journal} {Class. Quant. Grav.}\ }\textbf {\bibinfo
  {volume} {16}},\ \bibinfo {pages} {R159} (\bibinfo {year}
  {1999})}\BibitemShut {NoStop}%
\bibitem [{\citenamefont {Kokkotas}\ and\ \citenamefont
  {Schmidt}(1999)}]{Kokkotas:1999bd}%
  \BibitemOpen
  \bibfield  {author} {\bibinfo {author} {\bibfnamefont {K.~D.}\ \bibnamefont
  {Kokkotas}}\ and\ \bibinfo {author} {\bibfnamefont {B.~G.}\ \bibnamefont
  {Schmidt}},\ }\href {\doibase 10.12942/lrr-1999-2} {\bibfield  {journal}
  {\bibinfo  {journal} {Living Rev. Rel.}\ }\textbf {\bibinfo {volume} {2}},\
  \bibinfo {pages} {2} (\bibinfo {year} {1999})},\ \Eprint
  {http://arxiv.org/abs/gr-qc/9909058} {arXiv:gr-qc/9909058} \BibitemShut
  {NoStop}%
\bibitem [{\citenamefont {Berti}\ \emph {et~al.}(2025)\citenamefont {Berti}
  \emph {et~al.}}]{Berti:2025hly}%
  \BibitemOpen
  \bibfield  {author} {\bibinfo {author} {\bibfnamefont {E.}~\bibnamefont
  {Berti}} \emph {et~al.},\ }\href@noop {} {\  (\bibinfo {year} {2025})},\
  \Eprint {http://arxiv.org/abs/2505.23895} {arXiv:2505.23895 [gr-qc]}
  \BibitemShut {NoStop}%
\bibitem [{\citenamefont {Abbott}\ \emph {et~al.}(2016)\citenamefont {Abbott}
  \emph {et~al.}}]{LIGOScientific:2016lio}%
  \BibitemOpen
  \bibfield  {author} {\bibinfo {author} {\bibfnamefont {B.~P.}\ \bibnamefont
  {Abbott}} \emph {et~al.} (\bibinfo {collaboration} {LIGO Scientific,
  Virgo}),\ }\href {\doibase 10.1103/PhysRevLett.116.221101} {\bibfield
  {journal} {\bibinfo  {journal} {Phys. Rev. Lett.}\ }\textbf {\bibinfo
  {volume} {116}},\ \bibinfo {pages} {221101} (\bibinfo {year} {2016})},\
  \bibinfo {note} {[Erratum: Phys.Rev.Lett. 121, 129902 (2018)]},\ \Eprint
  {http://arxiv.org/abs/1602.03841} {arXiv:1602.03841 [gr-qc]} \BibitemShut
  {NoStop}%
\bibitem [{\citenamefont {Abbott}\ \emph
  {et~al.}(2021{\natexlab{a}})\citenamefont {Abbott} \emph
  {et~al.}}]{LIGOScientific:2020tif}%
  \BibitemOpen
  \bibfield  {author} {\bibinfo {author} {\bibfnamefont {R.}~\bibnamefont
  {Abbott}} \emph {et~al.} (\bibinfo {collaboration} {LIGO Scientific,
  Virgo}),\ }\href {\doibase 10.1103/PhysRevD.103.122002} {\bibfield  {journal}
  {\bibinfo  {journal} {Phys. Rev. D}\ }\textbf {\bibinfo {volume} {103}},\
  \bibinfo {pages} {122002} (\bibinfo {year} {2021}{\natexlab{a}})},\ \Eprint
  {http://arxiv.org/abs/2010.14529} {arXiv:2010.14529 [gr-qc]} \BibitemShut
  {NoStop}%
\bibitem [{\citenamefont {Abbott}\ \emph
  {et~al.}(2021{\natexlab{b}})\citenamefont {Abbott} \emph
  {et~al.}}]{LIGOScientific:2021sio}%
  \BibitemOpen
  \bibfield  {author} {\bibinfo {author} {\bibfnamefont {R.}~\bibnamefont
  {Abbott}} \emph {et~al.} (\bibinfo {collaboration} {LIGO Scientific, VIRGO,
  KAGRA}),\ }\href@noop {} {\  (\bibinfo {year} {2021}{\natexlab{b}})},\
  \Eprint {http://arxiv.org/abs/2112.06861} {arXiv:2112.06861 [gr-qc]}
  \BibitemShut {NoStop}%
\bibitem [{\citenamefont {Gong}\ \emph {et~al.}(2021)\citenamefont {Gong},
  \citenamefont {Luo},\ and\ \citenamefont {Wang}}]{Gong:2021gvw}%
  \BibitemOpen
  \bibfield  {author} {\bibinfo {author} {\bibfnamefont {Y.}~\bibnamefont
  {Gong}}, \bibinfo {author} {\bibfnamefont {J.}~\bibnamefont {Luo}}, \ and\
  \bibinfo {author} {\bibfnamefont {B.}~\bibnamefont {Wang}},\ }\href {\doibase
  10.1038/s41550-021-01480-3} {\bibfield  {journal} {\bibinfo  {journal}
  {Nature Astron.}\ }\textbf {\bibinfo {volume} {5}},\ \bibinfo {pages} {881}
  (\bibinfo {year} {2021})},\ \Eprint {http://arxiv.org/abs/2109.07442}
  {arXiv:2109.07442 [astro-ph.IM]} \BibitemShut {NoStop}%
\bibitem [{\citenamefont {Bigongiari}\ \emph {et~al.}(2025)\citenamefont
  {Bigongiari}, \citenamefont {Di~Giovanni},\ and\ \citenamefont
  {Losurdo}}]{Bigongiari:2025oyk}%
  \BibitemOpen
  \bibfield  {author} {\bibinfo {author} {\bibfnamefont {E.}~\bibnamefont
  {Bigongiari}}, \bibinfo {author} {\bibfnamefont {M.}~\bibnamefont
  {Di~Giovanni}}, \ and\ \bibinfo {author} {\bibfnamefont {G.}~\bibnamefont
  {Losurdo}},\ }\href@noop {} {\  (\bibinfo {year} {2025})},\ \Eprint
  {http://arxiv.org/abs/2509.25952} {arXiv:2509.25952 [gr-qc]} \BibitemShut
  {NoStop}%
\bibitem [{\citenamefont {Gossan}\ \emph {et~al.}(2012)\citenamefont {Gossan},
  \citenamefont {Veitch},\ and\ \citenamefont {Sathyaprakash}}]{Gossan:2011ha}%
  \BibitemOpen
  \bibfield  {author} {\bibinfo {author} {\bibfnamefont {S.}~\bibnamefont
  {Gossan}}, \bibinfo {author} {\bibfnamefont {J.}~\bibnamefont {Veitch}}, \
  and\ \bibinfo {author} {\bibfnamefont {B.~S.}\ \bibnamefont
  {Sathyaprakash}},\ }\href {\doibase 10.1103/PhysRevD.85.124056} {\bibfield
  {journal} {\bibinfo  {journal} {Phys. Rev. D}\ }\textbf {\bibinfo {volume}
  {85}},\ \bibinfo {pages} {124056} (\bibinfo {year} {2012})},\ \Eprint
  {http://arxiv.org/abs/1111.5819} {arXiv:1111.5819 [gr-qc]} \BibitemShut
  {NoStop}%
\bibitem [{\citenamefont {Meidam}\ \emph {et~al.}(2014)\citenamefont {Meidam},
  \citenamefont {Agathos}, \citenamefont {Van Den~Broeck}, \citenamefont
  {Veitch},\ and\ \citenamefont {Sathyaprakash}}]{Meidam:2014jpa}%
  \BibitemOpen
  \bibfield  {author} {\bibinfo {author} {\bibfnamefont {J.}~\bibnamefont
  {Meidam}}, \bibinfo {author} {\bibfnamefont {M.}~\bibnamefont {Agathos}},
  \bibinfo {author} {\bibfnamefont {C.}~\bibnamefont {Van Den~Broeck}},
  \bibinfo {author} {\bibfnamefont {J.}~\bibnamefont {Veitch}}, \ and\ \bibinfo
  {author} {\bibfnamefont {B.~S.}\ \bibnamefont {Sathyaprakash}},\ }\href
  {\doibase 10.1103/PhysRevD.90.064009} {\bibfield  {journal} {\bibinfo
  {journal} {Phys. Rev. D}\ }\textbf {\bibinfo {volume} {90}},\ \bibinfo
  {pages} {064009} (\bibinfo {year} {2014})},\ \Eprint
  {http://arxiv.org/abs/1406.3201} {arXiv:1406.3201 [gr-qc]} \BibitemShut
  {NoStop}%
\bibitem [{\citenamefont {Carullo}\ \emph {et~al.}(2018)\citenamefont {Carullo}
  \emph {et~al.}}]{Carullo:2018sfu}%
  \BibitemOpen
  \bibfield  {author} {\bibinfo {author} {\bibfnamefont {G.}~\bibnamefont
  {Carullo}} \emph {et~al.},\ }\href {\doibase 10.1103/PhysRevD.98.104020}
  {\bibfield  {journal} {\bibinfo  {journal} {Phys. Rev. D}\ }\textbf {\bibinfo
  {volume} {98}},\ \bibinfo {pages} {104020} (\bibinfo {year} {2018})},\
  \Eprint {http://arxiv.org/abs/1805.04760} {arXiv:1805.04760 [gr-qc]}
  \BibitemShut {NoStop}%
\bibitem [{\citenamefont {Abbott}\ \emph
  {et~al.}(2019{\natexlab{a}})\citenamefont {Abbott} \emph
  {et~al.}}]{LIGOScientific:2018dkp}%
  \BibitemOpen
  \bibfield  {author} {\bibinfo {author} {\bibfnamefont {B.~P.}\ \bibnamefont
  {Abbott}} \emph {et~al.} (\bibinfo {collaboration} {LIGO Scientific,
  Virgo}),\ }\href {\doibase 10.1103/PhysRevLett.123.011102} {\bibfield
  {journal} {\bibinfo  {journal} {Phys. Rev. Lett.}\ }\textbf {\bibinfo
  {volume} {123}},\ \bibinfo {pages} {011102} (\bibinfo {year}
  {2019}{\natexlab{a}})},\ \Eprint {http://arxiv.org/abs/1811.00364}
  {arXiv:1811.00364 [gr-qc]} \BibitemShut {NoStop}%
\bibitem [{\citenamefont {Abbott}\ \emph
  {et~al.}(2019{\natexlab{b}})\citenamefont {Abbott} \emph
  {et~al.}}]{LIGOScientific:2019fpa}%
  \BibitemOpen
  \bibfield  {author} {\bibinfo {author} {\bibfnamefont {B.~P.}\ \bibnamefont
  {Abbott}} \emph {et~al.} (\bibinfo {collaboration} {LIGO Scientific,
  Virgo}),\ }\href {\doibase 10.1103/PhysRevD.100.104036} {\bibfield  {journal}
  {\bibinfo  {journal} {Phys. Rev. D}\ }\textbf {\bibinfo {volume} {100}},\
  \bibinfo {pages} {104036} (\bibinfo {year} {2019}{\natexlab{b}})},\ \Eprint
  {http://arxiv.org/abs/1903.04467} {arXiv:1903.04467 [gr-qc]} \BibitemShut
  {NoStop}%
\bibitem [{\citenamefont {Yunes}\ \emph {et~al.}(2024)\citenamefont {Yunes},
  \citenamefont {Siemens},\ and\ \citenamefont {Yagi}}]{Yunes:2024lzm}%
  \BibitemOpen
  \bibfield  {author} {\bibinfo {author} {\bibfnamefont {N.}~\bibnamefont
  {Yunes}}, \bibinfo {author} {\bibfnamefont {X.}~\bibnamefont {Siemens}}, \
  and\ \bibinfo {author} {\bibfnamefont {K.}~\bibnamefont {Yagi}},\ }\href@noop
  {} {\  (\bibinfo {year} {2024})},\ \Eprint {http://arxiv.org/abs/2408.05240}
  {arXiv:2408.05240 [gr-qc]} \BibitemShut {NoStop}%
\bibitem [{\citenamefont {Tattersall}(2020)}]{Tattersall:2019nmh}%
  \BibitemOpen
  \bibfield  {author} {\bibinfo {author} {\bibfnamefont {O.~J.}\ \bibnamefont
  {Tattersall}},\ }\href {\doibase 10.1088/1361-6382/ab839b} {\bibfield
  {journal} {\bibinfo  {journal} {Class. Quant. Grav.}\ }\textbf {\bibinfo
  {volume} {37}},\ \bibinfo {pages} {115007} (\bibinfo {year} {2020})},\
  \Eprint {http://arxiv.org/abs/1911.07593} {arXiv:1911.07593 [gr-qc]}
  \BibitemShut {NoStop}%
\bibitem [{\citenamefont {de~Rham}\ \emph {et~al.}(2020)\citenamefont
  {de~Rham}, \citenamefont {Francfort},\ and\ \citenamefont
  {Zhang}}]{deRham:2020ejn}%
  \BibitemOpen
  \bibfield  {author} {\bibinfo {author} {\bibfnamefont {C.}~\bibnamefont
  {de~Rham}}, \bibinfo {author} {\bibfnamefont {J.}~\bibnamefont {Francfort}},
  \ and\ \bibinfo {author} {\bibfnamefont {J.}~\bibnamefont {Zhang}},\ }\href
  {\doibase 10.1103/PhysRevD.102.024079} {\bibfield  {journal} {\bibinfo
  {journal} {Phys. Rev. D}\ }\textbf {\bibinfo {volume} {102}},\ \bibinfo
  {pages} {024079} (\bibinfo {year} {2020})},\ \Eprint
  {http://arxiv.org/abs/2005.13923} {arXiv:2005.13923 [hep-th]} \BibitemShut
  {NoStop}%
\bibitem [{\citenamefont {Roussille}(2022)}]{Roussille:2022vfa}%
  \BibitemOpen
  \bibfield  {author} {\bibinfo {author} {\bibfnamefont {H.}~\bibnamefont
  {Roussille}},\ }\emph {\bibinfo {title} {{Black hole perturbations in
  modified gravity theories}}},\ \href@noop {} {Ph.D. thesis},\ \bibinfo
  {school} {Diderot U., Paris} (\bibinfo {year} {2022}),\ \Eprint
  {http://arxiv.org/abs/2211.01103} {arXiv:2211.01103 [gr-qc]} \BibitemShut
  {NoStop}%
\bibitem [{\citenamefont {Sirera}\ and\ \citenamefont
  {Noller}(2023)}]{Sirera:2023pbs}%
  \BibitemOpen
  \bibfield  {author} {\bibinfo {author} {\bibfnamefont {S.}~\bibnamefont
  {Sirera}}\ and\ \bibinfo {author} {\bibfnamefont {J.}~\bibnamefont
  {Noller}},\ }\href {\doibase 10.1103/PhysRevD.107.124054} {\bibfield
  {journal} {\bibinfo  {journal} {Phys. Rev. D}\ }\textbf {\bibinfo {volume}
  {107}},\ \bibinfo {pages} {124054} (\bibinfo {year} {2023})},\ \Eprint
  {http://arxiv.org/abs/2301.10272} {arXiv:2301.10272 [gr-qc]} \BibitemShut
  {NoStop}%
\bibitem [{\citenamefont {Mukohyama}\ \emph {et~al.}(2023)\citenamefont
  {Mukohyama}, \citenamefont {Takahashi}, \citenamefont {Tomikawa},\ and\
  \citenamefont {Yingcharoenrat}}]{Mukohyama:2023xyf}%
  \BibitemOpen
  \bibfield  {author} {\bibinfo {author} {\bibfnamefont {S.}~\bibnamefont
  {Mukohyama}}, \bibinfo {author} {\bibfnamefont {K.}~\bibnamefont
  {Takahashi}}, \bibinfo {author} {\bibfnamefont {K.}~\bibnamefont {Tomikawa}},
  \ and\ \bibinfo {author} {\bibfnamefont {V.}~\bibnamefont {Yingcharoenrat}},\
  }\href {\doibase 10.1088/1475-7516/2023/07/050} {\bibfield  {journal}
  {\bibinfo  {journal} {JCAP}\ }\textbf {\bibinfo {volume} {07}},\ \bibinfo
  {pages} {050} (\bibinfo {year} {2023})},\ \Eprint
  {http://arxiv.org/abs/2304.14304} {arXiv:2304.14304 [gr-qc]} \BibitemShut
  {NoStop}%
\bibitem [{\citenamefont {Leaver}(1985)}]{Leaver:1985ax}%
  \BibitemOpen
  \bibfield  {author} {\bibinfo {author} {\bibfnamefont {E.~W.}\ \bibnamefont
  {Leaver}},\ }\href {\doibase 10.1098/rspa.1985.0119} {\bibfield  {journal}
  {\bibinfo  {journal} {Proc. Roy. Soc. Lond. A}\ }\textbf {\bibinfo {volume}
  {402}},\ \bibinfo {pages} {285} (\bibinfo {year} {1985})}\BibitemShut
  {NoStop}%
\bibitem [{\citenamefont {Cardoso}\ \emph {et~al.}(2019)\citenamefont
  {Cardoso}, \citenamefont {Kimura}, \citenamefont {Maselli}, \citenamefont
  {Berti}, \citenamefont {Macedo},\ and\ \citenamefont
  {McManus}}]{Cardoso:2019mqo}%
  \BibitemOpen
  \bibfield  {author} {\bibinfo {author} {\bibfnamefont {V.}~\bibnamefont
  {Cardoso}}, \bibinfo {author} {\bibfnamefont {M.}~\bibnamefont {Kimura}},
  \bibinfo {author} {\bibfnamefont {A.}~\bibnamefont {Maselli}}, \bibinfo
  {author} {\bibfnamefont {E.}~\bibnamefont {Berti}}, \bibinfo {author}
  {\bibfnamefont {C.~F.~B.}\ \bibnamefont {Macedo}}, \ and\ \bibinfo {author}
  {\bibfnamefont {R.}~\bibnamefont {McManus}},\ }\href {\doibase
  10.1103/PhysRevD.99.104077} {\bibfield  {journal} {\bibinfo  {journal} {Phys.
  Rev. D}\ }\textbf {\bibinfo {volume} {99}},\ \bibinfo {pages} {104077}
  (\bibinfo {year} {2019})},\ \Eprint {http://arxiv.org/abs/1901.01265}
  {arXiv:1901.01265 [gr-qc]} \BibitemShut {NoStop}%
\bibitem [{\citenamefont {Hatsuda}\ and\ \citenamefont
  {Kimura}(2020)}]{Hatsuda:2020egs}%
  \BibitemOpen
  \bibfield  {author} {\bibinfo {author} {\bibfnamefont {Y.}~\bibnamefont
  {Hatsuda}}\ and\ \bibinfo {author} {\bibfnamefont {M.}~\bibnamefont
  {Kimura}},\ }\href {\doibase 10.1103/PhysRevD.102.044032} {\bibfield
  {journal} {\bibinfo  {journal} {Phys. Rev. D}\ }\textbf {\bibinfo {volume}
  {102}},\ \bibinfo {pages} {044032} (\bibinfo {year} {2020})},\ \Eprint
  {http://arxiv.org/abs/2006.15496} {arXiv:2006.15496 [gr-qc]} \BibitemShut
  {NoStop}%
\bibitem [{\citenamefont {V{\"o}lkel}\ \emph {et~al.}(2022)\citenamefont
  {V{\"o}lkel}, \citenamefont {Franchini},\ and\ \citenamefont
  {Barausse}}]{Volkel:2022aca}%
  \BibitemOpen
  \bibfield  {author} {\bibinfo {author} {\bibfnamefont {S.~H.}\ \bibnamefont
  {V{\"o}lkel}}, \bibinfo {author} {\bibfnamefont {N.}~\bibnamefont
  {Franchini}}, \ and\ \bibinfo {author} {\bibfnamefont {E.}~\bibnamefont
  {Barausse}},\ }\href {\doibase 10.1103/PhysRevD.105.084046} {\bibfield
  {journal} {\bibinfo  {journal} {Phys. Rev. D}\ }\textbf {\bibinfo {volume}
  {105}},\ \bibinfo {pages} {084046} (\bibinfo {year} {2022})},\ \Eprint
  {http://arxiv.org/abs/2202.08655} {arXiv:2202.08655 [gr-qc]} \BibitemShut
  {NoStop}%
\bibitem [{\citenamefont {McManus}\ \emph {et~al.}(2019)\citenamefont
  {McManus}, \citenamefont {Berti}, \citenamefont {Macedo}, \citenamefont
  {Kimura}, \citenamefont {Maselli},\ and\ \citenamefont
  {Cardoso}}]{McManus:2019ulj}%
  \BibitemOpen
  \bibfield  {author} {\bibinfo {author} {\bibfnamefont {R.}~\bibnamefont
  {McManus}}, \bibinfo {author} {\bibfnamefont {E.}~\bibnamefont {Berti}},
  \bibinfo {author} {\bibfnamefont {C.~F.~B.}\ \bibnamefont {Macedo}}, \bibinfo
  {author} {\bibfnamefont {M.}~\bibnamefont {Kimura}}, \bibinfo {author}
  {\bibfnamefont {A.}~\bibnamefont {Maselli}}, \ and\ \bibinfo {author}
  {\bibfnamefont {V.}~\bibnamefont {Cardoso}},\ }\href {\doibase
  10.1103/PhysRevD.100.044061} {\bibfield  {journal} {\bibinfo  {journal}
  {Phys. Rev. D}\ }\textbf {\bibinfo {volume} {100}},\ \bibinfo {pages}
  {044061} (\bibinfo {year} {2019})},\ \Eprint
  {http://arxiv.org/abs/1906.05155} {arXiv:1906.05155 [gr-qc]} \BibitemShut
  {NoStop}%
\bibitem [{\citenamefont {Hirano}\ \emph {et~al.}(2024)\citenamefont {Hirano},
  \citenamefont {Kimura}, \citenamefont {Yamaguchi},\ and\ \citenamefont
  {Zhang}}]{Hirano:2024fgp}%
  \BibitemOpen
  \bibfield  {author} {\bibinfo {author} {\bibfnamefont {S.}~\bibnamefont
  {Hirano}}, \bibinfo {author} {\bibfnamefont {M.}~\bibnamefont {Kimura}},
  \bibinfo {author} {\bibfnamefont {M.}~\bibnamefont {Yamaguchi}}, \ and\
  \bibinfo {author} {\bibfnamefont {J.}~\bibnamefont {Zhang}},\ }\href
  {\doibase 10.1103/PhysRevD.110.024015} {\bibfield  {journal} {\bibinfo
  {journal} {Phys. Rev. D}\ }\textbf {\bibinfo {volume} {110}},\ \bibinfo
  {pages} {024015} (\bibinfo {year} {2024})},\ \Eprint
  {http://arxiv.org/abs/2404.09672} {arXiv:2404.09672 [gr-qc]} \BibitemShut
  {NoStop}%
\bibitem [{\citenamefont {Thomopoulos}\ \emph {et~al.}(2025)\citenamefont
  {Thomopoulos}, \citenamefont {V{\"o}lkel},\ and\ \citenamefont
  {Pfeiffer}}]{Thomopoulos:2025nuf}%
  \BibitemOpen
  \bibfield  {author} {\bibinfo {author} {\bibfnamefont {S.}~\bibnamefont
  {Thomopoulos}}, \bibinfo {author} {\bibfnamefont {S.~H.}\ \bibnamefont
  {V{\"o}lkel}}, \ and\ \bibinfo {author} {\bibfnamefont {H.~P.}\ \bibnamefont
  {Pfeiffer}},\ }\href {\doibase 10.1103/xtzl-lyn6} {\bibfield  {journal}
  {\bibinfo  {journal} {Phys. Rev. D}\ }\textbf {\bibinfo {volume} {112}},\
  \bibinfo {pages} {064054} (\bibinfo {year} {2025})},\ \Eprint
  {http://arxiv.org/abs/2504.17848} {arXiv:2504.17848 [gr-qc]} \BibitemShut
  {NoStop}%
\bibitem [{\citenamefont {Cano}\ \emph {et~al.}(2024)\citenamefont {Cano},
  \citenamefont {Capuano}, \citenamefont {Franchini}, \citenamefont {Maenaut},\
  and\ \citenamefont {V{\"o}lkel}}]{Cano:2024jkd}%
  \BibitemOpen
  \bibfield  {author} {\bibinfo {author} {\bibfnamefont {P.~A.}\ \bibnamefont
  {Cano}}, \bibinfo {author} {\bibfnamefont {L.}~\bibnamefont {Capuano}},
  \bibinfo {author} {\bibfnamefont {N.}~\bibnamefont {Franchini}}, \bibinfo
  {author} {\bibfnamefont {S.}~\bibnamefont {Maenaut}}, \ and\ \bibinfo
  {author} {\bibfnamefont {S.~H.}\ \bibnamefont {V{\"o}lkel}},\ }\href
  {\doibase 10.1103/PhysRevD.110.104007} {\bibfield  {journal} {\bibinfo
  {journal} {Phys. Rev. D}\ }\textbf {\bibinfo {volume} {110}},\ \bibinfo
  {pages} {104007} (\bibinfo {year} {2024})},\ \Eprint
  {http://arxiv.org/abs/2407.15947} {arXiv:2407.15947 [gr-qc]} \BibitemShut
  {NoStop}%
\bibitem [{\citenamefont {Jaramillo}\ \emph {et~al.}(2021)\citenamefont
  {Jaramillo}, \citenamefont {Panosso~Macedo},\ and\ \citenamefont
  {Al~Sheikh}}]{Jaramillo:2020tuu}%
  \BibitemOpen
  \bibfield  {author} {\bibinfo {author} {\bibfnamefont {J.~L.}\ \bibnamefont
  {Jaramillo}}, \bibinfo {author} {\bibfnamefont {R.}~\bibnamefont
  {Panosso~Macedo}}, \ and\ \bibinfo {author} {\bibfnamefont {L.}~\bibnamefont
  {Al~Sheikh}},\ }\href {\doibase 10.1103/PhysRevX.11.031003} {\bibfield
  {journal} {\bibinfo  {journal} {Phys. Rev. X}\ }\textbf {\bibinfo {volume}
  {11}},\ \bibinfo {pages} {031003} (\bibinfo {year} {2021})},\ \Eprint
  {http://arxiv.org/abs/2004.06434} {arXiv:2004.06434 [gr-qc]} \BibitemShut
  {NoStop}%
\bibitem [{\citenamefont {Cai}\ \emph {et~al.}(2025)\citenamefont {Cai},
  \citenamefont {Cao}, \citenamefont {Chen}, \citenamefont {Guo}, \citenamefont
  {Wu},\ and\ \citenamefont {Zhou}}]{Cai:2025irl}%
  \BibitemOpen
  \bibfield  {author} {\bibinfo {author} {\bibfnamefont {R.-G.}\ \bibnamefont
  {Cai}}, \bibinfo {author} {\bibfnamefont {L.-M.}\ \bibnamefont {Cao}},
  \bibinfo {author} {\bibfnamefont {J.-N.}\ \bibnamefont {Chen}}, \bibinfo
  {author} {\bibfnamefont {Z.-K.}\ \bibnamefont {Guo}}, \bibinfo {author}
  {\bibfnamefont {L.-B.}\ \bibnamefont {Wu}}, \ and\ \bibinfo {author}
  {\bibfnamefont {Y.-S.}\ \bibnamefont {Zhou}},\ }\href {\doibase
  10.1103/PhysRevD.111.084011} {\bibfield  {journal} {\bibinfo  {journal}
  {Phys. Rev. D}\ }\textbf {\bibinfo {volume} {111}},\ \bibinfo {pages}
  {084011} (\bibinfo {year} {2025})},\ \Eprint
  {http://arxiv.org/abs/2501.02522} {arXiv:2501.02522 [gr-qc]} \BibitemShut
  {NoStop}%
\bibitem [{\citenamefont {Jansen}(2017)}]{Jansen:2017oag}%
  \BibitemOpen
  \bibfield  {author} {\bibinfo {author} {\bibfnamefont {A.}~\bibnamefont
  {Jansen}},\ }\href {\doibase 10.1140/epjp/i2017-11825-9} {\bibfield
  {journal} {\bibinfo  {journal} {Eur. Phys. J. Plus}\ }\textbf {\bibinfo
  {volume} {132}},\ \bibinfo {pages} {546} (\bibinfo {year} {2017})},\ \Eprint
  {http://arxiv.org/abs/1709.09178} {arXiv:1709.09178 [gr-qc]} \BibitemShut
  {NoStop}%
\bibitem [{\citenamefont {Ripley}(2022)}]{Ripley:2022ypi}%
  \BibitemOpen
  \bibfield  {author} {\bibinfo {author} {\bibfnamefont {J.~L.}\ \bibnamefont
  {Ripley}},\ }\href {\doibase 10.1088/1361-6382/ac776d} {\bibfield  {journal}
  {\bibinfo  {journal} {Class. Quant. Grav.}\ }\textbf {\bibinfo {volume}
  {39}},\ \bibinfo {pages} {145009} (\bibinfo {year} {2022})},\ \Eprint
  {http://arxiv.org/abs/2202.03837} {arXiv:2202.03837 [gr-qc]} \BibitemShut
  {NoStop}%
\bibitem [{\citenamefont {Panosso~Macedo}\ and\ \citenamefont
  {Zenginoglu}(2024)}]{PanossoMacedo:2024nkw}%
  \BibitemOpen
  \bibfield  {author} {\bibinfo {author} {\bibfnamefont {R.}~\bibnamefont
  {Panosso~Macedo}}\ and\ \bibinfo {author} {\bibfnamefont {A.}~\bibnamefont
  {Zenginoglu}},\ }\href {\doibase 10.3389/fphy.2024.1497601} {\bibfield
  {journal} {\bibinfo  {journal} {Front. in Phys.}\ }\textbf {\bibinfo {volume}
  {12}},\ \bibinfo {pages} {1497601} (\bibinfo {year} {2024})},\ \Eprint
  {http://arxiv.org/abs/2409.11478} {arXiv:2409.11478 [gr-qc]} \BibitemShut
  {NoStop}%
\bibitem [{\citenamefont {Xiong}\ and\ \citenamefont
  {Li}(2024)}]{Xiong:2024urw}%
  \BibitemOpen
  \bibfield  {author} {\bibinfo {author} {\bibfnamefont {W.}~\bibnamefont
  {Xiong}}\ and\ \bibinfo {author} {\bibfnamefont {P.-C.}\ \bibnamefont {Li}},\
  }\href@noop {} {\  (\bibinfo {year} {2024})},\ \Eprint
  {http://arxiv.org/abs/2411.19069} {arXiv:2411.19069 [gr-qc]} \BibitemShut
  {NoStop}%
\bibitem [{\citenamefont {Bl{\'a}zquez-Salcedo}\ \emph
  {et~al.}(2024)\citenamefont {Bl{\'a}zquez-Salcedo}, \citenamefont {Khoo},
  \citenamefont {Kunz},\ and\ \citenamefont
  {Gonz{\'a}lez-Romero}}]{Blazquez-Salcedo:2023hwg}%
  \BibitemOpen
  \bibfield  {author} {\bibinfo {author} {\bibfnamefont {J.~L.}\ \bibnamefont
  {Bl{\'a}zquez-Salcedo}}, \bibinfo {author} {\bibfnamefont {F.~S.}\
  \bibnamefont {Khoo}}, \bibinfo {author} {\bibfnamefont {J.}~\bibnamefont
  {Kunz}}, \ and\ \bibinfo {author} {\bibfnamefont {L.~M.}\ \bibnamefont
  {Gonz{\'a}lez-Romero}},\ }\href {\doibase 10.1103/PhysRevD.109.064028}
  {\bibfield  {journal} {\bibinfo  {journal} {Phys. Rev. D}\ }\textbf {\bibinfo
  {volume} {109}},\ \bibinfo {pages} {064028} (\bibinfo {year} {2024})},\
  \Eprint {http://arxiv.org/abs/2312.10754} {arXiv:2312.10754 [gr-qc]}
  \BibitemShut {NoStop}%
\bibitem [{\citenamefont {Chung}\ \emph {et~al.}(2024)\citenamefont {Chung},
  \citenamefont {Wagle},\ and\ \citenamefont {Yunes}}]{Chung:2023wkd}%
  \BibitemOpen
  \bibfield  {author} {\bibinfo {author} {\bibfnamefont {A.~K.-W.}\
  \bibnamefont {Chung}}, \bibinfo {author} {\bibfnamefont {P.}~\bibnamefont
  {Wagle}}, \ and\ \bibinfo {author} {\bibfnamefont {N.}~\bibnamefont
  {Yunes}},\ }\href {\doibase 10.1103/PhysRevD.109.044072} {\bibfield
  {journal} {\bibinfo  {journal} {Phys. Rev. D}\ }\textbf {\bibinfo {volume}
  {109}},\ \bibinfo {pages} {044072} (\bibinfo {year} {2024})},\ \Eprint
  {http://arxiv.org/abs/2312.08435} {arXiv:2312.08435 [gr-qc]} \BibitemShut
  {NoStop}%
\bibitem [{\citenamefont {Bl{\'a}zquez-Salcedo}\ \emph
  {et~al.}(2025)\citenamefont {Bl{\'a}zquez-Salcedo}, \citenamefont {Khoo},
  \citenamefont {Kleihaus},\ and\ \citenamefont
  {Kunz}}]{Blazquez-Salcedo:2024oek}%
  \BibitemOpen
  \bibfield  {author} {\bibinfo {author} {\bibfnamefont {J.~L.}\ \bibnamefont
  {Bl{\'a}zquez-Salcedo}}, \bibinfo {author} {\bibfnamefont {F.~S.}\
  \bibnamefont {Khoo}}, \bibinfo {author} {\bibfnamefont {B.}~\bibnamefont
  {Kleihaus}}, \ and\ \bibinfo {author} {\bibfnamefont {J.}~\bibnamefont
  {Kunz}},\ }\href {\doibase 10.1103/PhysRevD.111.L021505} {\bibfield
  {journal} {\bibinfo  {journal} {Phys. Rev. D}\ }\textbf {\bibinfo {volume}
  {111}},\ \bibinfo {pages} {L021505} (\bibinfo {year} {2025})},\ \Eprint
  {http://arxiv.org/abs/2407.20760} {arXiv:2407.20760 [gr-qc]} \BibitemShut
  {NoStop}%
\bibitem [{\citenamefont {Chung}\ and\ \citenamefont
  {Yunes}(2024{\natexlab{a}})}]{Chung:2024ira}%
  \BibitemOpen
  \bibfield  {author} {\bibinfo {author} {\bibfnamefont {A.~K.-W.}\
  \bibnamefont {Chung}}\ and\ \bibinfo {author} {\bibfnamefont
  {N.}~\bibnamefont {Yunes}},\ }\href {\doibase 10.1103/PhysRevLett.133.181401}
  {\bibfield  {journal} {\bibinfo  {journal} {Phys. Rev. Lett.}\ }\textbf
  {\bibinfo {volume} {133}},\ \bibinfo {pages} {181401} (\bibinfo {year}
  {2024}{\natexlab{a}})},\ \Eprint {http://arxiv.org/abs/2405.12280}
  {arXiv:2405.12280 [gr-qc]} \BibitemShut {NoStop}%
\bibitem [{\citenamefont {Chung}\ and\ \citenamefont
  {Yunes}(2024{\natexlab{b}})}]{Chung:2024vaf}%
  \BibitemOpen
  \bibfield  {author} {\bibinfo {author} {\bibfnamefont {A.~K.-W.}\
  \bibnamefont {Chung}}\ and\ \bibinfo {author} {\bibfnamefont
  {N.}~\bibnamefont {Yunes}},\ }\href {\doibase 10.1103/PhysRevD.110.064019}
  {\bibfield  {journal} {\bibinfo  {journal} {Phys. Rev. D}\ }\textbf {\bibinfo
  {volume} {110}},\ \bibinfo {pages} {064019} (\bibinfo {year}
  {2024}{\natexlab{b}})},\ \Eprint {http://arxiv.org/abs/2406.11986}
  {arXiv:2406.11986 [gr-qc]} \BibitemShut {NoStop}%
\bibitem [{\citenamefont {Cheung}\ \emph {et~al.}(2022)\citenamefont {Cheung},
  \citenamefont {Destounis}, \citenamefont {Macedo}, \citenamefont {Berti},\
  and\ \citenamefont {Cardoso}}]{Cheung:2021bol}%
  \BibitemOpen
  \bibfield  {author} {\bibinfo {author} {\bibfnamefont {M.~H.-Y.}\
  \bibnamefont {Cheung}}, \bibinfo {author} {\bibfnamefont {K.}~\bibnamefont
  {Destounis}}, \bibinfo {author} {\bibfnamefont {R.~P.}\ \bibnamefont
  {Macedo}}, \bibinfo {author} {\bibfnamefont {E.}~\bibnamefont {Berti}}, \
  and\ \bibinfo {author} {\bibfnamefont {V.}~\bibnamefont {Cardoso}},\ }\href
  {\doibase 10.1103/PhysRevLett.128.111103} {\bibfield  {journal} {\bibinfo
  {journal} {Phys. Rev. Lett.}\ }\textbf {\bibinfo {volume} {128}},\ \bibinfo
  {pages} {111103} (\bibinfo {year} {2022})},\ \Eprint
  {http://arxiv.org/abs/2111.05415} {arXiv:2111.05415 [gr-qc]} \BibitemShut
  {NoStop}%
\bibitem [{\citenamefont {Berti}\ \emph {et~al.}(2022)\citenamefont {Berti},
  \citenamefont {Cardoso}, \citenamefont {Cheung}, \citenamefont {Di~Filippo},
  \citenamefont {Duque}, \citenamefont {Martens},\ and\ \citenamefont
  {Mukohyama}}]{Berti:2022xfj}%
  \BibitemOpen
  \bibfield  {author} {\bibinfo {author} {\bibfnamefont {E.}~\bibnamefont
  {Berti}}, \bibinfo {author} {\bibfnamefont {V.}~\bibnamefont {Cardoso}},
  \bibinfo {author} {\bibfnamefont {M.~H.-Y.}\ \bibnamefont {Cheung}}, \bibinfo
  {author} {\bibfnamefont {F.}~\bibnamefont {Di~Filippo}}, \bibinfo {author}
  {\bibfnamefont {F.}~\bibnamefont {Duque}}, \bibinfo {author} {\bibfnamefont
  {P.}~\bibnamefont {Martens}}, \ and\ \bibinfo {author} {\bibfnamefont
  {S.}~\bibnamefont {Mukohyama}},\ }\href {\doibase
  10.1103/PhysRevD.106.084011} {\bibfield  {journal} {\bibinfo  {journal}
  {Phys. Rev. D}\ }\textbf {\bibinfo {volume} {106}},\ \bibinfo {pages}
  {084011} (\bibinfo {year} {2022})},\ \Eprint
  {http://arxiv.org/abs/2205.08547} {arXiv:2205.08547 [gr-qc]} \BibitemShut
  {NoStop}%
\bibitem [{\citenamefont {Courty}\ \emph {et~al.}(2023)\citenamefont {Courty},
  \citenamefont {Destounis},\ and\ \citenamefont {Pani}}]{Courty:2023rxk}%
  \BibitemOpen
  \bibfield  {author} {\bibinfo {author} {\bibfnamefont {A.}~\bibnamefont
  {Courty}}, \bibinfo {author} {\bibfnamefont {K.}~\bibnamefont {Destounis}}, \
  and\ \bibinfo {author} {\bibfnamefont {P.}~\bibnamefont {Pani}},\ }\href
  {\doibase 10.1103/PhysRevD.108.104027} {\bibfield  {journal} {\bibinfo
  {journal} {Phys. Rev. D}\ }\textbf {\bibinfo {volume} {108}},\ \bibinfo
  {pages} {104027} (\bibinfo {year} {2023})},\ \Eprint
  {http://arxiv.org/abs/2307.11155} {arXiv:2307.11155 [gr-qc]} \BibitemShut
  {NoStop}%
\bibitem [{\citenamefont {Yang}\ \emph {et~al.}(2024)\citenamefont {Yang},
  \citenamefont {Mai}, \citenamefont {Yang}, \citenamefont {Shao},\ and\
  \citenamefont {Berti}}]{Yang:2024vor}%
  \BibitemOpen
  \bibfield  {author} {\bibinfo {author} {\bibfnamefont {Y.}~\bibnamefont
  {Yang}}, \bibinfo {author} {\bibfnamefont {Z.-F.}\ \bibnamefont {Mai}},
  \bibinfo {author} {\bibfnamefont {R.-Q.}\ \bibnamefont {Yang}}, \bibinfo
  {author} {\bibfnamefont {L.}~\bibnamefont {Shao}}, \ and\ \bibinfo {author}
  {\bibfnamefont {E.}~\bibnamefont {Berti}},\ }\href {\doibase
  10.1103/PhysRevD.110.084018} {\bibfield  {journal} {\bibinfo  {journal}
  {Phys. Rev. D}\ }\textbf {\bibinfo {volume} {110}},\ \bibinfo {pages}
  {084018} (\bibinfo {year} {2024})},\ \Eprint
  {http://arxiv.org/abs/2407.20131} {arXiv:2407.20131 [gr-qc]} \BibitemShut
  {NoStop}%
\bibitem [{\citenamefont {Cardoso}\ \emph {et~al.}(2024)\citenamefont
  {Cardoso}, \citenamefont {Kastha},\ and\ \citenamefont
  {Panosso~Macedo}}]{Cardoso:2024mrw}%
  \BibitemOpen
  \bibfield  {author} {\bibinfo {author} {\bibfnamefont {V.}~\bibnamefont
  {Cardoso}}, \bibinfo {author} {\bibfnamefont {S.}~\bibnamefont {Kastha}}, \
  and\ \bibinfo {author} {\bibfnamefont {R.}~\bibnamefont {Panosso~Macedo}},\
  }\href {\doibase 10.1103/PhysRevD.110.024016} {\bibfield  {journal} {\bibinfo
   {journal} {Phys. Rev. D}\ }\textbf {\bibinfo {volume} {110}},\ \bibinfo
  {pages} {024016} (\bibinfo {year} {2024})},\ \Eprint
  {http://arxiv.org/abs/2404.01374} {arXiv:2404.01374 [gr-qc]} \BibitemShut
  {NoStop}%
\bibitem [{\citenamefont {Ianniccari}\ \emph {et~al.}(2024)\citenamefont
  {Ianniccari}, \citenamefont {Iovino}, \citenamefont {Kehagias}, \citenamefont
  {Pani}, \citenamefont {Perna}, \citenamefont {Perrone},\ and\ \citenamefont
  {Riotto}}]{Ianniccari:2024ysv}%
  \BibitemOpen
  \bibfield  {author} {\bibinfo {author} {\bibfnamefont {A.}~\bibnamefont
  {Ianniccari}}, \bibinfo {author} {\bibfnamefont {A.~J.}\ \bibnamefont
  {Iovino}}, \bibinfo {author} {\bibfnamefont {A.}~\bibnamefont {Kehagias}},
  \bibinfo {author} {\bibfnamefont {P.}~\bibnamefont {Pani}}, \bibinfo {author}
  {\bibfnamefont {G.}~\bibnamefont {Perna}}, \bibinfo {author} {\bibfnamefont
  {D.}~\bibnamefont {Perrone}}, \ and\ \bibinfo {author} {\bibfnamefont
  {A.}~\bibnamefont {Riotto}},\ }\href {\doibase
  10.1103/PhysRevLett.133.211401} {\bibfield  {journal} {\bibinfo  {journal}
  {Phys. Rev. Lett.}\ }\textbf {\bibinfo {volume} {133}},\ \bibinfo {pages}
  {211401} (\bibinfo {year} {2024})},\ \Eprint
  {http://arxiv.org/abs/2407.20144} {arXiv:2407.20144 [gr-qc]} \BibitemShut
  {NoStop}%
\bibitem [{\citenamefont {Malato~Corr{\^e}a}\ \emph {et~al.}(2025)\citenamefont
  {Malato~Corr{\^e}a}, \citenamefont {Macedo}, \citenamefont {Panosso~Macedo},\
  and\ \citenamefont {Oliveira}}]{MalatoCorrea:2025iuc}%
  \BibitemOpen
  \bibfield  {author} {\bibinfo {author} {\bibfnamefont {M.}~\bibnamefont
  {Malato~Corr{\^e}a}}, \bibinfo {author} {\bibfnamefont {C.~F.~B.}\
  \bibnamefont {Macedo}}, \bibinfo {author} {\bibfnamefont {R.}~\bibnamefont
  {Panosso~Macedo}}, \ and\ \bibinfo {author} {\bibfnamefont {L.~A.}\
  \bibnamefont {Oliveira}},\ }\href {\doibase 10.1103/78ht-dn36} {\bibfield
  {journal} {\bibinfo  {journal} {Phys. Rev. D}\ }\textbf {\bibinfo {volume}
  {112}},\ \bibinfo {pages} {024036} (\bibinfo {year} {2025})},\ \Eprint
  {http://arxiv.org/abs/2504.00107} {arXiv:2504.00107 [gr-qc]} \BibitemShut
  {NoStop}%
\bibitem [{\citenamefont {Shen}\ \emph
  {et~al.}(2025{\natexlab{a}})\citenamefont {Shen}, \citenamefont {Li},
  \citenamefont {Daghigh}, \citenamefont {Morey}, \citenamefont {Green},
  \citenamefont {Qian},\ and\ \citenamefont {Yue}}]{Shen:2025nsl}%
  \BibitemOpen
  \bibfield  {author} {\bibinfo {author} {\bibfnamefont {S.-F.}\ \bibnamefont
  {Shen}}, \bibinfo {author} {\bibfnamefont {G.-R.}\ \bibnamefont {Li}},
  \bibinfo {author} {\bibfnamefont {R.~G.}\ \bibnamefont {Daghigh}}, \bibinfo
  {author} {\bibfnamefont {J.~C.}\ \bibnamefont {Morey}}, \bibinfo {author}
  {\bibfnamefont {M.~D.}\ \bibnamefont {Green}}, \bibinfo {author}
  {\bibfnamefont {W.-L.}\ \bibnamefont {Qian}}, \ and\ \bibinfo {author}
  {\bibfnamefont {R.-H.}\ \bibnamefont {Yue}},\ }\href@noop {} {\  (\bibinfo
  {year} {2025}{\natexlab{a}})},\ \Eprint {http://arxiv.org/abs/2509.23372}
  {arXiv:2509.23372 [gr-qc]} \BibitemShut {NoStop}%
\bibitem [{\citenamefont {Mai}\ and\ \citenamefont {Yang}(2025)}]{Mai:2025cva}%
  \BibitemOpen
  \bibfield  {author} {\bibinfo {author} {\bibfnamefont {Z.-F.}\ \bibnamefont
  {Mai}}\ and\ \bibinfo {author} {\bibfnamefont {R.-Q.}\ \bibnamefont {Yang}},\
  }\href@noop {} {\  (\bibinfo {year} {2025})},\ \Eprint
  {http://arxiv.org/abs/2506.07562} {arXiv:2506.07562 [gr-qc]} \BibitemShut
  {NoStop}%
\bibitem [{\citenamefont {Bini}\ \emph {et~al.}(2002)\citenamefont {Bini},
  \citenamefont {Cherubini}, \citenamefont {Jantzen},\ and\ \citenamefont
  {Ruffini}}]{Bini:2002jx}%
  \BibitemOpen
  \bibfield  {author} {\bibinfo {author} {\bibfnamefont {D.}~\bibnamefont
  {Bini}}, \bibinfo {author} {\bibfnamefont {C.}~\bibnamefont {Cherubini}},
  \bibinfo {author} {\bibfnamefont {R.~T.}\ \bibnamefont {Jantzen}}, \ and\
  \bibinfo {author} {\bibfnamefont {R.~J.}\ \bibnamefont {Ruffini}},\ }\href
  {\doibase 10.1143/PTP.107.967} {\bibfield  {journal} {\bibinfo  {journal}
  {Prog. Theor. Phys.}\ }\textbf {\bibinfo {volume} {107}},\ \bibinfo {pages}
  {967} (\bibinfo {year} {2002})},\ \Eprint
  {http://arxiv.org/abs/gr-qc/0203069} {arXiv:gr-qc/0203069} \BibitemShut
  {NoStop}%
\bibitem [{\citenamefont {Konoplya}\ and\ \citenamefont
  {Zhidenko}(2011)}]{Konoplya:2011qq}%
  \BibitemOpen
  \bibfield  {author} {\bibinfo {author} {\bibfnamefont {R.~A.}\ \bibnamefont
  {Konoplya}}\ and\ \bibinfo {author} {\bibfnamefont {A.}~\bibnamefont
  {Zhidenko}},\ }\href {\doibase 10.1103/RevModPhys.83.793} {\bibfield
  {journal} {\bibinfo  {journal} {Rev. Mod. Phys.}\ }\textbf {\bibinfo {volume}
  {83}},\ \bibinfo {pages} {793} (\bibinfo {year} {2011})},\ \Eprint
  {http://arxiv.org/abs/1102.4014} {arXiv:1102.4014 [gr-qc]} \BibitemShut
  {NoStop}%
\bibitem [{\citenamefont {Chen}\ \emph {et~al.}(2025)\citenamefont {Chen},
  \citenamefont {Guo},\ and\ \citenamefont {Wu}}]{Chen:2025wfi}%
  \BibitemOpen
  \bibfield  {author} {\bibinfo {author} {\bibfnamefont {J.-N.}\ \bibnamefont
  {Chen}}, \bibinfo {author} {\bibfnamefont {Z.-K.}\ \bibnamefont {Guo}}, \
  and\ \bibinfo {author} {\bibfnamefont {L.-B.}\ \bibnamefont {Wu}},\
  }\href@noop {} {\  (\bibinfo {year} {2025})},\ \Eprint
  {http://arxiv.org/abs/2510.27320} {arXiv:2510.27320 [gr-qc]} \BibitemShut
  {NoStop}%
\bibitem [{\citenamefont {Assaad}\ and\ \citenamefont
  {Panosso~Macedo}(2025)}]{Assaad:2025nbv}%
  \BibitemOpen
  \bibfield  {author} {\bibinfo {author} {\bibfnamefont {J.}~\bibnamefont
  {Assaad}}\ and\ \bibinfo {author} {\bibfnamefont {R.}~\bibnamefont
  {Panosso~Macedo}},\ }\href@noop {} {\  (\bibinfo {year} {2025})},\ \Eprint
  {http://arxiv.org/abs/2506.04326} {arXiv:2506.04326 [gr-qc]} \BibitemShut
  {NoStop}%
\bibitem [{\citenamefont {Ansorg}\ and\ \citenamefont
  {Panosso~Macedo}(2016)}]{Ansorg:2016ztf}%
  \BibitemOpen
  \bibfield  {author} {\bibinfo {author} {\bibfnamefont {M.}~\bibnamefont
  {Ansorg}}\ and\ \bibinfo {author} {\bibfnamefont {R.}~\bibnamefont
  {Panosso~Macedo}},\ }\href {\doibase 10.1103/PhysRevD.93.124016} {\bibfield
  {journal} {\bibinfo  {journal} {Phys. Rev. D}\ }\textbf {\bibinfo {volume}
  {93}},\ \bibinfo {pages} {124016} (\bibinfo {year} {2016})},\ \Eprint
  {http://arxiv.org/abs/1604.02261} {arXiv:1604.02261 [gr-qc]} \BibitemShut
  {NoStop}%
\bibitem [{\citenamefont {Panosso~Macedo}\ \emph {et~al.}(2018)\citenamefont
  {Panosso~Macedo}, \citenamefont {Jaramillo},\ and\ \citenamefont
  {Ansorg}}]{PanossoMacedo:2018hab}%
  \BibitemOpen
  \bibfield  {author} {\bibinfo {author} {\bibfnamefont {R.}~\bibnamefont
  {Panosso~Macedo}}, \bibinfo {author} {\bibfnamefont {J.~L.}\ \bibnamefont
  {Jaramillo}}, \ and\ \bibinfo {author} {\bibfnamefont {M.}~\bibnamefont
  {Ansorg}},\ }\href {\doibase 10.1103/PhysRevD.98.124005} {\bibfield
  {journal} {\bibinfo  {journal} {Phys. Rev. D}\ }\textbf {\bibinfo {volume}
  {98}},\ \bibinfo {pages} {124005} (\bibinfo {year} {2018})},\ \Eprint
  {http://arxiv.org/abs/1809.02837} {arXiv:1809.02837 [gr-qc]} \BibitemShut
  {NoStop}%
\bibitem [{\citenamefont {Panosso~Macedo}(2020)}]{PanossoMacedo:2019npm}%
  \BibitemOpen
  \bibfield  {author} {\bibinfo {author} {\bibfnamefont {R.}~\bibnamefont
  {Panosso~Macedo}},\ }\href {\doibase 10.1088/1361-6382/ab6e3e} {\bibfield
  {journal} {\bibinfo  {journal} {Class. Quant. Grav.}\ }\textbf {\bibinfo
  {volume} {37}},\ \bibinfo {pages} {065019} (\bibinfo {year} {2020})},\
  \Eprint {http://arxiv.org/abs/1910.13452} {arXiv:1910.13452 [gr-qc]}
  \BibitemShut {NoStop}%
\bibitem [{\citenamefont {Panosso~Macedo}(2024)}]{PanossoMacedo:2023qzp}%
  \BibitemOpen
  \bibfield  {author} {\bibinfo {author} {\bibfnamefont {R.}~\bibnamefont
  {Panosso~Macedo}},\ }\href {\doibase 10.1098/rsta.2023.0046} {\bibfield
  {journal} {\bibinfo  {journal} {Phil. Trans. Roy. Soc. Lond. A}\ }\textbf
  {\bibinfo {volume} {382}},\ \bibinfo {pages} {20230046} (\bibinfo {year}
  {2024})},\ \Eprint {http://arxiv.org/abs/2307.15735} {arXiv:2307.15735
  [gr-qc]} \BibitemShut {NoStop}%
\bibitem [{\citenamefont {Zhou}\ and\ \citenamefont
  {Panosso~Macedo}(2025)}]{Zhou:2025xta}%
  \BibitemOpen
  \bibfield  {author} {\bibinfo {author} {\bibfnamefont {Y.}~\bibnamefont
  {Zhou}}\ and\ \bibinfo {author} {\bibfnamefont {R.}~\bibnamefont
  {Panosso~Macedo}},\ }\href {\doibase 10.1103/n948-jnfh} {\bibfield  {journal}
  {\bibinfo  {journal} {Phys. Rev. D}\ }\textbf {\bibinfo {volume} {112}},\
  \bibinfo {pages} {084063} (\bibinfo {year} {2025})},\ \Eprint
  {http://arxiv.org/abs/2507.05370} {arXiv:2507.05370 [gr-qc]} \BibitemShut
  {NoStop}%
\bibitem [{\citenamefont {Destounis}\ \emph {et~al.}(2021)\citenamefont
  {Destounis}, \citenamefont {Macedo}, \citenamefont {Berti}, \citenamefont
  {Cardoso},\ and\ \citenamefont {Jaramillo}}]{Destounis:2021lum}%
  \BibitemOpen
  \bibfield  {author} {\bibinfo {author} {\bibfnamefont {K.}~\bibnamefont
  {Destounis}}, \bibinfo {author} {\bibfnamefont {R.~P.}\ \bibnamefont
  {Macedo}}, \bibinfo {author} {\bibfnamefont {E.}~\bibnamefont {Berti}},
  \bibinfo {author} {\bibfnamefont {V.}~\bibnamefont {Cardoso}}, \ and\
  \bibinfo {author} {\bibfnamefont {J.~L.}\ \bibnamefont {Jaramillo}},\ }\href
  {\doibase 10.1103/PhysRevD.104.084091} {\bibfield  {journal} {\bibinfo
  {journal} {Phys. Rev. D}\ }\textbf {\bibinfo {volume} {104}},\ \bibinfo
  {pages} {084091} (\bibinfo {year} {2021})},\ \Eprint
  {http://arxiv.org/abs/2107.09673} {arXiv:2107.09673 [gr-qc]} \BibitemShut
  {NoStop}%
\bibitem [{\citenamefont {Cao}\ \emph {et~al.}(2024)\citenamefont {Cao},
  \citenamefont {Chen}, \citenamefont {Wu}, \citenamefont {Xie},\ and\
  \citenamefont {Zhou}}]{Cao:2024oud}%
  \BibitemOpen
  \bibfield  {author} {\bibinfo {author} {\bibfnamefont {L.-M.}\ \bibnamefont
  {Cao}}, \bibinfo {author} {\bibfnamefont {J.-N.}\ \bibnamefont {Chen}},
  \bibinfo {author} {\bibfnamefont {L.-B.}\ \bibnamefont {Wu}}, \bibinfo
  {author} {\bibfnamefont {L.}~\bibnamefont {Xie}}, \ and\ \bibinfo {author}
  {\bibfnamefont {Y.-S.}\ \bibnamefont {Zhou}},\ }\href {\doibase
  10.1007/s11433-024-2435-5} {\bibfield  {journal} {\bibinfo  {journal} {Sci.
  China Phys. Mech. Astron.}\ }\textbf {\bibinfo {volume} {67}},\ \bibinfo
  {pages} {100412} (\bibinfo {year} {2024})},\ \Eprint
  {http://arxiv.org/abs/2401.09907} {arXiv:2401.09907 [gr-qc]} \BibitemShut
  {NoStop}%
\bibitem [{\citenamefont {Wu}\ \emph {et~al.}(2025)\citenamefont {Wu},
  \citenamefont {Cai},\ and\ \citenamefont {Xie}}]{Wu:2024ldo}%
  \BibitemOpen
  \bibfield  {author} {\bibinfo {author} {\bibfnamefont {L.-B.}\ \bibnamefont
  {Wu}}, \bibinfo {author} {\bibfnamefont {R.-G.}\ \bibnamefont {Cai}}, \ and\
  \bibinfo {author} {\bibfnamefont {L.}~\bibnamefont {Xie}},\ }\href {\doibase
  10.1103/PhysRevD.111.044066} {\bibfield  {journal} {\bibinfo  {journal}
  {Phys. Rev. D}\ }\textbf {\bibinfo {volume} {111}},\ \bibinfo {pages}
  {044066} (\bibinfo {year} {2025})},\ \Eprint
  {http://arxiv.org/abs/2411.07734} {arXiv:2411.07734 [gr-qc]} \BibitemShut
  {NoStop}%
\bibitem [{\citenamefont {Cao}\ \emph {et~al.}(2025{\natexlab{a}})\citenamefont
  {Cao}, \citenamefont {Wu},\ and\ \citenamefont {Zhou}}]{Cao:2024sot}%
  \BibitemOpen
  \bibfield  {author} {\bibinfo {author} {\bibfnamefont {L.-M.}\ \bibnamefont
  {Cao}}, \bibinfo {author} {\bibfnamefont {L.-B.}\ \bibnamefont {Wu}}, \ and\
  \bibinfo {author} {\bibfnamefont {Y.-S.}\ \bibnamefont {Zhou}},\ }\href
  {\doibase 10.1007/s11433-025-2714-7} {\bibfield  {journal} {\bibinfo
  {journal} {Sci. China Phys. Mech. Astron.}\ }\textbf {\bibinfo {volume}
  {68}},\ \bibinfo {pages} {100411} (\bibinfo {year} {2025}{\natexlab{a}})},\
  \Eprint {http://arxiv.org/abs/2412.21092} {arXiv:2412.21092 [gr-qc]}
  \BibitemShut {NoStop}%
\bibitem [{\citenamefont {Yang}\ \emph {et~al.}(2025)\citenamefont {Yang},
  \citenamefont {Wu}, \citenamefont {Kuang},\ and\ \citenamefont
  {Qian}}]{Yang:2025hqk}%
  \BibitemOpen
  \bibfield  {author} {\bibinfo {author} {\bibfnamefont {Z.-H.}\ \bibnamefont
  {Yang}}, \bibinfo {author} {\bibfnamefont {L.-B.}\ \bibnamefont {Wu}},
  \bibinfo {author} {\bibfnamefont {X.-M.}\ \bibnamefont {Kuang}}, \ and\
  \bibinfo {author} {\bibfnamefont {W.-L.}\ \bibnamefont {Qian}},\ }\href@noop
  {} {\  (\bibinfo {year} {2025})},\ \Eprint {http://arxiv.org/abs/2510.02033}
  {arXiv:2510.02033 [gr-qc]} \BibitemShut {NoStop}%
\bibitem [{\citenamefont {Cao}\ \emph {et~al.}(2025{\natexlab{b}})\citenamefont
  {Cao}, \citenamefont {Ji}, \citenamefont {Wu},\ and\ \citenamefont
  {Zhou}}]{Cao:2025qws}%
  \BibitemOpen
  \bibfield  {author} {\bibinfo {author} {\bibfnamefont {L.-M.}\ \bibnamefont
  {Cao}}, \bibinfo {author} {\bibfnamefont {M.-F.}\ \bibnamefont {Ji}},
  \bibinfo {author} {\bibfnamefont {L.-B.}\ \bibnamefont {Wu}}, \ and\ \bibinfo
  {author} {\bibfnamefont {Y.-S.}\ \bibnamefont {Zhou}},\ }\href@noop {} {\
  (\bibinfo {year} {2025}{\natexlab{b}})},\ \Eprint
  {http://arxiv.org/abs/2508.13894} {arXiv:2508.13894 [gr-qc]} \BibitemShut
  {NoStop}%
\bibitem [{\citenamefont {Shen}\ \emph
  {et~al.}(2025{\natexlab{b}})\citenamefont {Shen}, \citenamefont {Li},
  \citenamefont {Kuang}, \citenamefont {Qian}, \citenamefont {Daghigh},
  \citenamefont {Morey}, \citenamefont {Green},\ and\ \citenamefont
  {Yue}}]{Shen:2025nbq}%
  \BibitemOpen
  \bibfield  {author} {\bibinfo {author} {\bibfnamefont {S.-F.}\ \bibnamefont
  {Shen}}, \bibinfo {author} {\bibfnamefont {G.-R.}\ \bibnamefont {Li}},
  \bibinfo {author} {\bibfnamefont {X.-M.}\ \bibnamefont {Kuang}}, \bibinfo
  {author} {\bibfnamefont {W.-L.}\ \bibnamefont {Qian}}, \bibinfo {author}
  {\bibfnamefont {R.~G.}\ \bibnamefont {Daghigh}}, \bibinfo {author}
  {\bibfnamefont {J.~C.}\ \bibnamefont {Morey}}, \bibinfo {author}
  {\bibfnamefont {M.~D.}\ \bibnamefont {Green}}, \ and\ \bibinfo {author}
  {\bibfnamefont {R.-H.}\ \bibnamefont {Yue}},\ }\href@noop {} {\  (\bibinfo
  {year} {2025}{\natexlab{b}})},\ \Eprint {http://arxiv.org/abs/2508.09031}
  {arXiv:2508.09031 [gr-qc]} \BibitemShut {NoStop}%
\bibitem [{\citenamefont {Chen}\ \emph {et~al.}(2024)\citenamefont {Chen},
  \citenamefont {Wu},\ and\ \citenamefont {Guo}}]{Chen:2024mon}%
  \BibitemOpen
  \bibfield  {author} {\bibinfo {author} {\bibfnamefont {J.-N.}\ \bibnamefont
  {Chen}}, \bibinfo {author} {\bibfnamefont {L.-B.}\ \bibnamefont {Wu}}, \ and\
  \bibinfo {author} {\bibfnamefont {Z.-K.}\ \bibnamefont {Guo}},\ }\href
  {\doibase 10.1088/1361-6382/ad89a1} {\bibfield  {journal} {\bibinfo
  {journal} {Class. Quant. Grav.}\ }\textbf {\bibinfo {volume} {41}},\ \bibinfo
  {pages} {235015} (\bibinfo {year} {2024})},\ \Eprint
  {http://arxiv.org/abs/2407.03907} {arXiv:2407.03907 [gr-qc]} \BibitemShut
  {NoStop}%
\bibitem [{\citenamefont {Destounis}\ \emph {et~al.}(2024)\citenamefont
  {Destounis}, \citenamefont {Boyanov},\ and\ \citenamefont
  {Panosso~Macedo}}]{Destounis:2023nmb}%
  \BibitemOpen
  \bibfield  {author} {\bibinfo {author} {\bibfnamefont {K.}~\bibnamefont
  {Destounis}}, \bibinfo {author} {\bibfnamefont {V.}~\bibnamefont {Boyanov}},
  \ and\ \bibinfo {author} {\bibfnamefont {R.}~\bibnamefont {Panosso~Macedo}},\
  }\href {\doibase 10.1103/PhysRevD.109.044023} {\bibfield  {journal} {\bibinfo
   {journal} {Phys. Rev. D}\ }\textbf {\bibinfo {volume} {109}},\ \bibinfo
  {pages} {044023} (\bibinfo {year} {2024})},\ \Eprint
  {http://arxiv.org/abs/2312.11630} {arXiv:2312.11630 [gr-qc]} \BibitemShut
  {NoStop}%
\bibitem [{\citenamefont {Cownden}\ \emph {et~al.}(2024)\citenamefont
  {Cownden}, \citenamefont {Pantelidou},\ and\ \citenamefont
  {Zilh{\~a}o}}]{Cownden:2023dam}%
  \BibitemOpen
  \bibfield  {author} {\bibinfo {author} {\bibfnamefont {B.}~\bibnamefont
  {Cownden}}, \bibinfo {author} {\bibfnamefont {C.}~\bibnamefont {Pantelidou}},
  \ and\ \bibinfo {author} {\bibfnamefont {M.}~\bibnamefont {Zilh{\~a}o}},\
  }\href {\doibase 10.1007/JHEP05(2024)202} {\bibfield  {journal} {\bibinfo
  {journal} {JHEP}\ }\textbf {\bibinfo {volume} {05}},\ \bibinfo {pages} {202}
  (\bibinfo {year} {2024})},\ \Eprint {http://arxiv.org/abs/2312.08352}
  {arXiv:2312.08352 [gr-qc]} \BibitemShut {NoStop}%
\bibitem [{\citenamefont {Boyanov}\ \emph {et~al.}(2024)\citenamefont
  {Boyanov}, \citenamefont {Cardoso}, \citenamefont {Destounis}, \citenamefont
  {Jaramillo},\ and\ \citenamefont {Panosso~Macedo}}]{Boyanov:2023qqf}%
  \BibitemOpen
  \bibfield  {author} {\bibinfo {author} {\bibfnamefont {V.}~\bibnamefont
  {Boyanov}}, \bibinfo {author} {\bibfnamefont {V.}~\bibnamefont {Cardoso}},
  \bibinfo {author} {\bibfnamefont {K.}~\bibnamefont {Destounis}}, \bibinfo
  {author} {\bibfnamefont {J.~L.}\ \bibnamefont {Jaramillo}}, \ and\ \bibinfo
  {author} {\bibfnamefont {R.}~\bibnamefont {Panosso~Macedo}},\ }\href
  {\doibase 10.1103/PhysRevD.109.064068} {\bibfield  {journal} {\bibinfo
  {journal} {Phys. Rev. D}\ }\textbf {\bibinfo {volume} {109}},\ \bibinfo
  {pages} {064068} (\bibinfo {year} {2024})},\ \Eprint
  {http://arxiv.org/abs/2312.11998} {arXiv:2312.11998 [gr-qc]} \BibitemShut
  {NoStop}%
\bibitem [{\citenamefont {Trefethen}(2000)}]{doi:10.1137/1.9780898719598}%
  \BibitemOpen
  \bibfield  {author} {\bibinfo {author} {\bibfnamefont {L.~N.}\ \bibnamefont
  {Trefethen}},\ }\href {\doibase 10.1137/1.9780898719598} {\emph {\bibinfo
  {title} {Spectral Methods in MATLAB}}}\ (\bibinfo  {publisher} {Society for
  Industrial and Applied Mathematics},\ \bibinfo {year} {2000})\BibitemShut
  {NoStop}%
\bibitem [{\citenamefont {Miguel}(2024)}]{Miguel:2023rzp}%
  \BibitemOpen
  \bibfield  {author} {\bibinfo {author} {\bibfnamefont {F.~S.}\ \bibnamefont
  {Miguel}},\ }\href {\doibase 10.1103/PhysRevD.109.104016} {\bibfield
  {journal} {\bibinfo  {journal} {Phys. Rev. D}\ }\textbf {\bibinfo {volume}
  {109}},\ \bibinfo {pages} {104016} (\bibinfo {year} {2024})},\ \Eprint
  {http://arxiv.org/abs/2308.03832} {arXiv:2308.03832 [gr-qc]} \BibitemShut
  {NoStop}%
\end{thebibliography}%
\bibliographystyle{apsrev4-1}

\end{document}